\documentclass[pra,aps,twocolumn,showpacs,nofootinbib]{revtex4-1}

\usepackage{savesym}
\usepackage{amsmath}
\savesymbol{iint}
\usepackage{txfonts}
\restoresymbol{TXF}{iint}

\usepackage{setspace}
\usepackage{psfrag}

\usepackage{bm,color}
\usepackage{amssymb,amsfonts}
\usepackage{setspace}

\usepackage[unicode]{hyperref}
\hypersetup{
    unicode=true,                                           
    a4paper=true,
    plainpages=false,
    pdffitwindow=true,                                      
    pdftitle={Number-conserving approaches to n-component Bose-Einstein condensates},                      
    pdfauthor={Peter Mason and Simon A. Gardiner},                            
    pdfsubject={},                                          
    pdfkeywords={Bose-Einstein, condensate, BEC},
    colorlinks=true,                                        
    linkcolor=blue,                                         
    citecolor=blue,                                         
    filecolor=blue,                                         
    urlcolor=blue                                           
}
\def\rmi{i}

\def\Pkd{\hat{\Psi}_k^{\dagger}(\bm{r})}
\def\Pk{\hat{\Psi}_k(\bm{r})}
\def\Pjd{\hat{\Psi}_j^{\dagger}(\bm{r})}
\def\Pj{\hat{\Psi}_j(\bm{r})}
\def\P1d{\hat{\Psi}_1^{\dagger}(\bm{r})}
\def\P1{\hat{\Psi}_1(\bm{r})}
\def\P2d{\hat{\Psi}_2^{\dagger}(\bm{r})}
\def\P2{\hat{\Psi}_2(\bm{r})}

\def\Pkddd{\hat{\Psi}_{k'}^{\dagger}(\bm{r}')}
\def\Pkdd{\hat{\Psi}_k^{\dagger}(\bm{r}')}
\def\Pkrd{\hat{\Psi}_k(\bm{r}')}

\def\P1dd{\hat{\Psi}_1^{\dagger}(\bm{r}')}
\def\P1d{\hat{\Psi}_1(\bm{r}')}
\def\P2dd{\hat{\Psi}_2^{\dagger}(\bm{r}')}
\def\P2d{\hat{\Psi}_2(\bm{r}')}

\def\Hk{H_{\text{sp}}^k(\bm{r},t)}

\def\H1{H_{\text{sp}}^1(\bm{r},t)}
\def\H2{H_{\text{sp}}^2(\bm{r},t)}
\def\Ujk{U_{jk}}
\def\Ukk{U_{kk}}

\def\U12{U_{12}}
\def\U21{U_{21}}

\def\U1{U_1}
\def\U2{U_2}

\def\skd1n{\sum_{k'=1}^n}
\def\sk1n{\sum_{k=1}^n}
\def\sj3k{\sum_{\substack{j=1\\j\neq k}}^3}

\def\sjkn{\sum_{\substack{j,k\\j<k}}^n}
\def\sjkdn{\sum_{\substack{j=1\\j\neq k'}}^n}

\def\sjk{\sum_{\substack{j=1\\j< k}}^n}
\def\sjdkdn{\sum_{\substack{j=1\\j\neq k}}^n}

\def\dr{\,d\bm{r}}

\def\drdd{\,d\bm{r}''}

\def\drrd{\delta(\bm{r}-\bm{r}')}

\def\Utjk{\tilde{U}_{jk}}
\def\Utkk{\tilde{U}_{kk}}

\def\Ut12{\tilde{U}_{12}}
\def\Ut21{\tilde{U}_{21}}

\def\Utkdkd{\tilde{U}_{k'k'}}

\def\Ut1{\tilde{U}_1}
\def\Ut2{\tilde{U}_2}

\def\npkd{N_{c_{p(k')}}}
\def\nhpk{\hat{N}_{c_{p(k)}}}
\def\nhpkd{\hat{N}_{c_{p(k')}}}

\def\npk{N_{c_{p(k)}}}
\def\np3k{N_{c_{3-k}}}
\def\nk{N_{c_{k}}}

\def\nhpj{\hat{N}_{c_{p(j)}}}
\def\npj{N_{c_{p(j)}}}

\def\Ltk{\tilde{\Lambda}_k(\bm{r},t)}
\def\Ltkdd{\tilde{\Lambda}_k^{\dagger}(\bm{r}',t)}
\def\Ltkddd{\tilde{\Lambda}_{k'}^{\dagger}(\bm{r}',t)}
\def\Ltkd{\tilde{\Lambda}_{k'}(\bm{r},t)}

\def\Lk{\tilde{\Lambda}_k(\bm{r})}
\def\Lkdag{\tilde{\Lambda}_k^{\dagger}(\bm{r})}

\def\Lkddd{\tilde{\Lambda}_{k'}^{\dagger}(\bm{r}')}
\def\Lkd{\tilde{\Lambda}_{k'}(\bm{r})}
\def\Lkdash{\tilde{\Lambda}_{k'}(\bm{r'})}

\def\Lj{\tilde{\Lambda}_j(\bm{r})}
\def\Ljdag{\tilde{\Lambda}_j^{\dagger}(\bm{r})}

\begin{document}

\title{Number-conserving approaches to $n$-component Bose--Einstein condensates}
\author {Peter Mason and Simon A. Gardiner}
\address{Joint Quantum Centre (JQC) Durham-Newcastle, Department of Physics, Durham University, Durham DH1 3LE, United Kingdom}
\date{\today}

\begin{abstract}
We develop the number-conserving approach that has previously been used in a single component Bose--Einstein condensed dilute atomic gas, to describe consistent coupled condensate and noncondensate number dynamics, to an $n$-component condensate. The resulting system of equations comprises, for each component, of a generalised Gross--Pitaevskii equation coupled to modified Bogoliubov--de Gennes equations. Lower-order approximations yield general formulations for multi-component Gross--Pitaevskii equations, and systems of multi-component Gross--Pitaevskii equations coupled to multi-component modified number-conserving Bogoliubov--de Gennes equations.  The analysis is left general, such that, in the $n$-component condensate, there may or may not be mutually coherent components. An expansion in powers of the ratio of noncondensate to condensate particle numbers for each coherent set is used to derive the self-consistent, second-order, dynamical equations of motion. The advantage of the analysis developed in this article is in its applications to dynamical instabilities that appear when two (or more) components are in conflict and where a significant noncondensed fraction of atoms is expected to appear.
\end{abstract}
\maketitle


\section{Introduction}
\label{intro}

Modern experimental apparatus allows a Bose--Einstein condensate consisting of a single species of atom to be realised at ultra-low temperatures (typically of the order of  nano-Kelvin \cite{bloch}).  Thermal effects are then frequently considered negligible, to the extent that a Hartree--Fock mean-field approach is often utilised \cite{ps,pet}. In this zero-temperature limit, the resultant dynamical description of the condensate is provided by the Gross--Pitaevskii (GP) equation (a cubic nonlinear Schr\"odinger equation), which propagates a classical field with a form reminiscent of a Schr\"{o}dinger wave function \cite{ps}. One of the basic assumptions in justifying this mean-field description is to assume that effectively all particles have Bose-condensed. Yet interatomic interactions between the individual atoms directly implies the existence of a small noncondensate fraction, for any finite total atom number, even at zero temperature. Such a noncondensate fraction can become non-negligible, particularly at finite temperature \cite{prou,gardbook}, or when there is a dynamical depletion of the condensate \cite{sag04,sag03,sag97,sag00,zhang1,zhang2,shep,reslen,mont,bg,bmg,gard2002}, such as occurs when the condensate exhibits non-equilibrium dynamics \cite{gardbook,bmg}.

There is an increasing interest in multi-component condensates, where in general the noncondensate fraction is likely to be a quantity of experimental significance.  It is by now fair to say that
single component Bose--Einstein condensates can be created relatively readily.  A large number of different alkali atoms (and many of their isotopes), as well as some non-alkali species, have been condensed \cite{bloch} and we understand well, both experimentally and theoretically, the effect of an applied magnetic field on the $s$-wave scattering length through Feshbach resonances \cite{inoyue,roberts,marte,blackley}.  When one considers the experimental realisation of a multi-component condensate, for instance of two-component condensates \cite{matthews,myatt,mod1,papp,durham}, then the situation becomes more involved. The scattering lengths must be determined for each atomic species pair, as has been accomplished for example for $^{85}$Rb-$^{87}$Rb \cite{papp}, $^{85}$Rb-$^{133}$Cs \cite{cho}, $^{87}$Rb-$^{133}$Cs \cite{pilch} or for $^{41}$Rb-$^{87}$K \cite{simoni} mixtures. Variation of these scattering lengths can lead to miscible or immiscible (phase separated) condensates, and a number of theoretical works have looked to establish the equilibrium density profiles \cite{ma,tsubota,m2013} and criteria for phase separation \cite{timm,wen,ole}. However, there may be a difference in the temperature at which the condensed components are initially held (if condensed separately), and there can easily be situations in which non-equilibrium dynamics are prevalent. A number of studies have reported the development of fundamental instabilities in two-component condensates, such as collective oscillations in colliding condensates \cite{madd,mod2}, Rayleigh--Taylor-type instabilities \cite{sasaki,gautam,kad}, Kelvin--Helmholtz-type instabilities \cite{tak,lundh}, counter-superflow instabilities \cite{itt,tit}, or crossovers between Kelvin--Helmholtz-type instabilities and counter-superflow instabilities \cite{stkts}. A separate line of studies has focused on ``exotic'' condensates, such as the spin-orbit condensate \cite{dalirev,am2013,achilleos}, the two- or three- component condensate with a Rabi coupling \cite{mattia,nitta}, spin-orbit together with Rabi coupling \cite{sala}, or dipole--dipole interactions in two-component condensates \cite{wilson}. However, while these studies are often driven by an applied external field such as a potential gradient or a rotation, they are, in general, carried out in the mean-field limit, i.e.\ formally they assume that there is no noncondensate fraction present in any of the components. 

Our primary interest in this article is thus to develop a consistent description of the dynamical interaction between condensed and noncondensed fractions of  multi-component Bose-condensed systems in order to facilitate an understanding of the dynamics present in the lead-up to instabilities. As such, our description of the condensate-noncondensate dynamics concentrates on situations where the origin of the noncondensate fraction is mainly dynamical, rather than thermal. This approach is particularly suited for a dynamical depletion of the condensate parts \cite{bmg}, however small thermal fractions can also be incorporated \cite{bmg}. As such, application of our results to study the effect of spin squeezing in finite temperature condensates is entirely possible \cite{sorensen,sin1,sin2}. The approach we take builds on the work of \cite{bg,bmg,gm}, who developed a self-consistent second-order number-conserving formalism for single component condensates, which itself owes its origin to the works of Gardiner \cite{g} and Castin \& Dum \cite{cd}. The starting point in any such number-conserving description is through the Penrose--Onsager criterion, in which a single-body density matrix is defined in terms of quantum field operators. The subsequent analysis then proceeds by splitting this quantum field operator into a condensate and a noncondensate part, which allows one, through an expansion in terms of powers of the (small) ratio of the noncondensate to condensate particle numbers, to develop a set of coupled equations, describing the condensate with a generalised Gross--Pitaevskii equation, and the noncondensate with modified Bogoliubov--de Gennes equations. The formulation that we develop in this article requires a non-trivial generalisation of this procedure. We must firstly define the single-body density matrix, but since our system contains in general $n$ components, this definition needs to be adapted to account for those components that are mutually coherent or incoherent. To separate the components into mutually coherent and incoherent sets is merely a formal bookkeeping; each set contains the components which are all coupled through the same one-body term, such as a Rabi coupling term (the elements are mutually coherent), but there are no coherences with respect to all the other sets. The resulting analysis collects the quantum field operators into distinct coherent sets, which is crucial for our partitioning of each operator into condensate and noncondensate parts, and in our definition of the expansion parameter. 

Through this expansion parameter we are able to derive a set of self-consistent second-order equations which comprise, for each component, a generalised Gross--Pitaevskii equation coupled to modified Bogoliubov--de Gennes equations. In the process of doing so, we also provide general derivations for multi-component time-dependent Gross--Pitaevskii equations, and multi-component equivalents to the non-self-consistent, but nonetheless useful system of the Gross--Piteavskii equation coupled to modified Bogoliubov--de Gennes equations.

Despite our concentration on a number-conserving approach, it is pertinent to remark on other possible theories that could be employed, particularly if one were to look to a multi-component condensate in which thermal effects were expected to play a large part in the dynamics. The number-conserving approach explicitly partitions the system into orthogonal condensate and noncondensate parts. In contrast one could employ a symmetry-breaking approach in which the $U(1)$ global phase symmetry is broken by describing the quantum field operator as a sum of a (c-number) finite expectation value and a fluctuation term around this expectation value. The expectation term is thus, in general, not orthogonal to the fluctuation term. We note that symmetry-breaking formulations can only conserve the mean particle number, as the grand canonical ensemble is required to give the field operator a finite expectation value. The Hartree--Fock--Bogoliubov--Popov or Zaremba--Nikuni--Griffin descriptions of the single component condensate are specific examples of symmetry-breaking approaches \cite{gardbook}, the former of which have recently been applied to two-component condensates \cite{lsp,roy}. They both rely on a perturbative expansion about a mean-field, which is philosophically similar to the number-conserving approach taken in this paper.

Another line of description, for example those that lead to a stochastic Gross--Pitaevskii equation, a projected Gross--Pitaevskii equation, or a stochastic projected Gross--Pitaevskii equation, relies on a c-field method whereby highly occupied modes of the system are described in terms of a classical field. At present, however, these methods are limited --- particularly, the latter two ``projected'' descriptions are only applicable in a high-temperature regime since they do not consider a quantum treatment of the pair-excitation process, a process that drives condensate depletion (see \cite{gardbook} for an overview).

A development of these theories to describe the multi-component condensate is certainly warranted, however, we are concerned with a self-consistent treatment of the number dynamics within a multi-component condensate. Our paper is thus organised as follows. Section \ref{form} begins by introducing the effective Hamiltonian and the $n$ quantum field operators and fluctuation operators (expansion parameters). Section \ref{pert} then presents an expansion of the effective Hamiltonian in terms of the expansion parameters, while in Sec.\ \ref{evolu} we derive the time-evolution equations of the particle operators and the fluctuation operators. Sections \ref{zero} and \ref{there} then proceed to derive the equations of motion that describe the condensate and noncondensate dynamics of the multi-component condensate. This is followed (Sec.\ \ref{egs}) by specific examples for a two- and three- (mutually coherent or incoherent) component condensate. Section \ref{concs} comprises the conclusions, and is followed by five technical appendices.

\section{Formulation}
\label{form}

\subsection{Overview of chosen formulation}
\label{form_1}

In this section we introduce the effective Hamiltonian that describes our $n$-component condensate. This Hamiltonian will be written down, and developed, in its most general form, i.e.\ to include the possibility for there to be coherent couplings between any of the components, such as a Rabi coupling or a synthetic gauge coupling. A consequence of including this generality into the Hamiltonian is the need to carefully establish a suitable notation in the subsequent formal development. This takes the form of a partitioning of the sample space of components into coherent subsets, and an associated mapping that takes any given collection of mutually coherent components into a specific subset.

Upon establishing this notation, we can write down the single-body density matrix for the $n$-component condensate. Our analysis relies on us establishing condensate and noncondensate representations for each component in order to track their dynamical evolution, and so we proceed by partitioning each of the field operators into a condensate and noncondensate part. We will then need to introduce an expansion parameter allowing us to develop a third-order effective Hamiltonian (established in Sec.\ \ref{pert}). In Sec.\ \ref{form_5} we define the fluctuation operators, equivalent to small expansion parameters, one for each component, that will be used throughout this paper in order to develop the self-consistent set of dynamical equations describing the $n$-component condensate. We will conclude this section by looking at two- and three-component condensate examples in order to clarify the preceding development of the general effective Hamiltonian.

\subsection{Effective Hamiltonian}
\label{form_2}

Our system consists of an $n$-component Bose--Einstein condensate described in terms of $n$ particle-field operators $\hat{\Psi}_k$ ($k=1,\dots,n$) that are subject to the usual bosonic commutation relations 
\begin{subequations}
	\begin{align}
	\left[\Pj,\Pkdd\right]&=\delta_{jk}\drrd,\\
	\left[\Pj,\Pkrd\right]&=\left[\Pjd,\Pkdd\right]=0,
	\end{align}
\end{subequations}
where the index of the particle-field operator may refer to different internal states of the same atomic species or to entirely different species of atom. Here $\delta_{jk}$ is the Kronecker delta. We consider the system to contain $N$ bosons that undergo pairwise interactions with bosons in the same component and with bosons in different components. As such, we replace the true interaction potentials $V^{\text{bin}}$ with energy-independent contact potentials (pseudopotentials), defined as
\begin{equation}
	V^{\text{bin}}_{jk}(\bm{r}-\bm{r}')=\Ujk\drrd,
\end{equation}
where, for a three-dimensional cold dilute Bose gas,
\begin{equation}
	\Ujk=\frac{2\pi\hbar^2a_{jk}}{M_{jk}},
\end{equation}
where $\hbar$ is Planck's constant, the $s$-wave scattering lengths are $a_{jk}$ and where $M_{jk}$ is the reduced mass, given by $M_{jk}^{-1}=M_j^{-1}+M_k^{-1}$ with $M_k$ the atomic mass of a boson in component $k$. We call the $U_{kk}$ the intracomponent coupling and the $U_{jk}$ ($j\neq k$) the intercomponent coupling. In general $a_{jk}=a_{kj}$ and so $U_{jk}=U_{kj}$. We note here that the local pseudopotential must be regularised (renormalised according to various quantities appearing in the subsequent development of the theory) in order to avoid ultraviolent divergencies \cite{gm,morgan}.

We restrict ourselves to considering only condensates where there are no spin-changing collisions, i.e.\ the magnetic sublevels must be resolved, for example through the application of a small bias magnetic field \cite{foot1}.  As such, the only interactions between different components that we consider here are density--density interactions or coherent couplings between internal atomic states. We can then write down the binary interaction effective Hamiltonian for the $n$-component condensate as
\begin{equation}
	\label{Ham}
	\begin{split}
	\hat{H}(t)=&\int\sk1n\Big[\Pkd\Hk\Pk\\
	&+\frac{\Ukk}{2}\Pkd\Pkd\Pk\Pk\Big]\dr\\
	&+\int\sum_{\substack{j,k\\j< k}}^n\Ujk\Pjd\Pkd\Pj\Pk\dr\\
	&+\int\sum_{\substack{j,k\\j\neq k}}^n\Pjd H_{\text{ob}}^{jk}(\bm{r},t)\Pk\dr.
	\end{split}
\end{equation}
The first term in (\ref{Ham}) contains kinetic and external potential energy terms, so that a typical single-particle Hamiltonian \cite{foot2} 
for component $k$ is given by
\begin{equation}
	\label{single}
	\Hk=-\frac{\hbar^2}{2M_k}\nabla^2+V_k(\bm{r},t)+\hbar \nu_{k},
\end{equation}
where $V_k(\bm{r},t)$ is an external potential (in general taken to be time dependent) applied to component $k$, and $\nu_{k}$ accounts for energy differences between different atomic internal states. The terms involving $U_{jk}$ in the above Hamiltonian [(\ref{Ham}), second and third terms] account for density--density (two-body) interactions within and between components. 

The final term in the Hamiltonian represents any coherent (one-body) coupling of atomic internal states; the precise form of $H_{\text{ob}}^{jk}(\bm{r},t)$ is not of concern for us in our subsequent treatment of this effective Hamiltonian, however a simple example might be 
\begin{equation}
	H_{\text{ob}}^{jk}(\bm{r},t)=\frac{\hbar}{2}\omega_{jk}\exp[\text{sgn}(j-k)i\theta_{jk}].
\end{equation}
This describes Rabi couplings, where the $\omega_{jk}$ denote the respective Rabi frequencies between the different component internal states and $\theta_{jk}$ accounts for any phase (both of which may in general be time-dependent), and the sign function is defined by
\begin{equation}
	\text{sgn}(j-k)=
	\begin{cases}
		+1\quad\text{if}\quad j>k,\\
		-1\quad\text{if}\quad j<k.
	\end{cases}
\end{equation}
Note that $\omega_{jk}=\omega_{kj}$ and $\theta_{jk}=\theta_{kj}$ so that the matrices $\Omega=(\omega_{jk})$ and $\Theta=(\theta_{jk})$ are symmetric (hence Hermitian). We note that, by definition, $\theta_{kk}=0$  always.   Another possibility would be to consider a synthetic gauge field, such as a spin-$1/2$ Rashba coupling \cite{am2013}, in which case one would also expect extra terms beyond those considered in (\ref{single}) to appear in $\Hk$.

\subsection{Mutually coherent and incoherent components}
\label{form_3}

The effective Hamiltonian has, for generality, included a term determining the internal coupling between all components --- manifested by the one-body term $H_{\text{ob}}^{jk}(\bm{r},t)$. However in the theory that we formulate, we do not impose coherence between any specific components. 
We will need to include a mechanism by which the precise nature of the condensate can be easily input into the system whilst leaving this precise nature unspecified. To this extent, we denote the sample space of component field operators as
$	\mathsf{\hat{\Psi}}=\{\hat{\Psi}_k\}$,
where $k\in [1,n]$. 
Now define a subset $p_i$ of $\mathsf{\hat{\Psi}}$ as a set whose elements are all coherent. 
We thus have $l$ ($l\in[1,n]$) subsets of $\mathsf{\hat{\Psi}}$, labelled $p_i$, such that
	\begin{align*}
		\bigcup_{i=1}^lp_i&=\mathsf{\hat{\Psi}}\quad\text{the sample space},\\
		\bigcap_{i=1}^lp_i&=\emptyset\quad\text{the empty set},
	\end{align*}
where $\bigcup$ represents a union of all subsets, and $\bigcap$ represents an intersection of all subsets, i.e.\ the $p_i$ are pairwise mutually exclusive and exhaustive for $\mathsf{\hat{\Psi}}$, forming a partition of $\mathsf{\hat{\Psi}}$. We define $|p_i|=m_i$ so that $\sum_{i=1}^lm_i=|\mathsf{\hat{\Psi}}|=n$. Note that if we were to choose an $n$-component condensate in which all components were mutually coherent then $l=1$ and $m_1=n$, or conversely if we were to choose an $n$-component condensate in which all components were mutually incoherent then $l=n$ and $m_i=1$ for all $i\in[1,n]$.  There are, in general, $C$ distinct ways to realise the subsets $p_{i}$.

At this stage we introduce a notation concerning the subset that each component belongs to. We choose component $k$, with field operator $\hat{\Psi}_k$, to be in some subset $p_i$, where we leave the choice of the index $i$ undetermined. Similarly, we say component $k'$ ($k'$ being different to $k$), with associated field operator $\hat{\Psi}_{k'}$, is in some subset $p_j$ ($j\in[1,n]$). The case in which $i=j$ corresponds to component $k$ and component $k'$ being in the same subset --- i.e.\ component $k$ and component $k'$ are mutually coherent components (and are necessarily the same isotope of an atomic species with different internal spin states). Conversely the case in which $i\neq j$ corresponds to component $k$ and component $k'$ being in different subsets, i.e.\ component $k$ and component $k'$ are mutually incoherent components (they are different atomic species, or different isotopes of the same atomic species, or even a mixture of different spin states which are not mutually coherent). To account for the structure of $\mathsf{\hat{\Psi}}$ we define $p(k)$ to be a mapping for component field $\hat{\Psi}_k$ to the subset $p_i$ containing component $k$, i.e.
\begin{equation}
	p(k): \hat{\Psi}_k\rightarrow p_i.
\end{equation}
The above is a formal way of stating that we order our component field operators into sets which have elements (component field operators) that are all mutually coherent to one another. In the following we will often make the equivalence $p_i\equiv i$ for convenience. 

\subsection{Density matrix and condensate or non-condensate parts}
\label{form_4}

Our analysis proceeds as in \cite{gm,cd} and we define a single-body density matrix, of form $\rho_{kk'}(\bm{r},\bm{r}',t)$, for the particles, given as $\rho_{kk'}(\bm{r},\bm{r}',t)=\langle\Pkddd\Pk\rangle.$
For our multi-component case, we choose to include a Kronecker delta-type term to account for the fact that components could be mutually coherent or incoherent. Thus we define the single-body density matrix by \begin{equation}
	\label{matrix}
	\rho_{kk'}(\bm{r},\bm{r}',t)=\left<\Pkddd\Pk\right>\delta_{k,k'}^p,
\end{equation}
where $\delta_{k,k'}^p$ is a Kronecker delta ``mapping'' term defined by \cite{foot3}
\begin{equation*}
	\delta_{k,k'}^p=
	\begin{cases}
		1\qquad{\text{mappings $p(k)$ and $p(k')$ are identical}},\\
		0\qquad{\text{mappings $p(k)$ and $p(k')$ are different}}.
	\end{cases}
\end{equation*}
This single-body density matrix is Hermitian, and so it can be decomposed into a complete set of eigenfunctions with related real eigenvalues. Since we suppose that each of the individual components is Bose--Einstein condensed, we are free to assume that each component has a single distinct ``large'' eigenfunction $\phi_k(\bm{r},t)$. Then each subset $p_i$ has a corresponding eigenvalue significantly larger than all the other eigenvalues associated with that subset. We define these eigenfunctions to have unit norm and thus write
\begin{equation}
	\label{N}
	\sum^{n}_{k'=1}\int\rho_{kk'}(\bm{r},\bm{r}',t)\phi_{k'}(\bm{r}',t)\dr'=N_{c_{p{(k)}}}(t)\phi_k(\bm{r},t),
\end{equation}
where $N_{c_{p{(k)}}}(t)$ is the eigenvalue associated with the subset containing component $k$.
We call the $\phi_k(\bm{r},t)$ the condensate parts, and similarly define noncondensate field operators $\delta\hat{\Psi}_k(\bm{r},t)$ for each component, such that the field operator 
${\Pk}$ 
is partitioned as \cite{foot4}
\begin{equation}
	\label{cond}
	\Pk=\hat{a}_{c_{p(k)}}(t)\phi_k(\bm{r},t)+\delta\hat{\Psi}_k(\bm{r},t).
\end{equation}
Here, the $\hat{a}_{c_{p(k)}}(t)$ are annihilation operators for particles in $\phi_k(\bm{r},t)$ with associated mapping $p(k)$, and $\delta\hat{\Psi}_k(\bm{r},t)$ is the part of the field operator 
$\Pk$ 
that is orthogonal to $\phi_k(\bm{r},t)$. As such, the $\hat{a}^{\dagger}_{c_{p(k)}}(t)$ are creation operators defined as
\begin{equation}
	\label{a}
	\hat{a}^{\dagger}_{c_{p(k)}}(t)=\sum^{n}_{k'=1}\delta_{k,k'}^p\int\hat{\Psi}^{\dagger}_{k'}(\bm{r})\phi_{k'}(\bm{r},t)\dr,
\end{equation}
and the noncondensate field operators are defined as
\begin{equation}
	\label{delta}
	\delta\hat{\Psi}_k(\bm{r},t)=\sum_{k'=1}^{n}\int Q_{kk'}(\bm{r},\bm{r}',t)\hat{\Psi}_{k'}(\bm{r}')\dr',
\end{equation}
where the projector $Q_{kk'}(\bm{r},\bm{r}',t)$ is defined as
\begin{equation}
	\label{qdeq}
	Q_{kk'}(\bm{r},\bm{r}',t)=\left[\delta_{kk'}\drrd-\phi_k(\bm{r},t)\phi_{k'}^*(\bm{r}',t)\right]\delta_{k,k'}^p.
\end{equation}
This means that the only nonzero commutation relations that involve $\hat{a}_{c_{p(k)}}(t)$ and $\delta\hat{\Psi}_k(\bm{r},t)$, and their Hermitian conjugates, are \cite{foot5}
\begin{subequations}
\begin{align}
	[\hat{a}_{c_{p(k)}}(t),\hat{a}^{\dagger}_{c_{p(k')}}(t)]&=\delta_{k,k'}^p,\\
	\label{qqq}
	\big[\delta\hat{\Psi}_k(\bm{r},t)
	,\delta\hat{\Psi}^{\dagger}_{k'}(\bm{r}',t)\big]
	&=Q_{kk'}(\bm{r},\bm{r}',t).
\end{align}
\label{commutator}
\end{subequations}

Introducing the single-body density matrix (\ref{matrix}) and the partition of the field operators into a condensate and noncondensate part [(\ref{N}) and (\ref{cond})] means that we can define ${\hat{N}_{c_{p(k)}}(t)}\equiv\hat{a}^{\dagger}_{c_{p(k)}}(t)\hat{a}_{c_{p(k)}}(t)$, from which it follows that
\begin{equation}
	\left<\hat{a}^{\dagger}_{c_{p(k)}}(t)\hat{a}_{c_{p(k)}}(t)\right>=\left<{\hat{N}_{c_{p(k)}}(t)}\right>
	 ={N_{c_{p(k)}}(t)}.
\end{equation}
It is then clear that the eigenvalue $N_{c_{p(k)}}(t)$ is the mean number of particles in the condensate part with associated mapping $p(k)$. We note that
\begin{equation}
	\label{equal0}
	\left<\hat{a}^{\dagger}_{c_{p(k)}}(t)\delta\hat{\Psi}_{k'}(\bm{r},t)\right>=0\qquad\forall{k,k'},
\end{equation}
stating that there are no simple coherences between the condensate part with mapping $p(k)$ and (any of) the noncondensate parts. For our system of $N$ bosonic atoms, we suppose that the total number of condensed atoms is $N_c(t)$, so that $\sum_{i=1}^l N_{c_{p{(i)}}}(t)=N_c(t)$. It follows that the total number of noncondensed atoms is $N_t(t)=N-N_c(t)$. At this point one can define the number of noncondensed atoms: Let $N_{c_{p{(k)}}}(t)$ be the number of noncondensed atoms associated with the subset containing component $k$. Then $\sum_{i=1}^l N_{t_{p{(i)}}}(t)=N_t(t)$. By assumption we have, for all $k$ and $k'$, $N_{c_{p{(k)}}}(t)\gg N_{t_{p{(k')}}}(t)$.

\subsection{Fluctuation operators}
\label{form_5}

We choose to perform a perturbation expansion on the effective Hamilton (\ref{Ham}) using ``fluctuation'' operators \cite{gm,ole} defined as
\begin{equation}
	\label{fluc}
	\Ltk=\frac{1}{\sqrt{N_{c_{p(k)}}(t)}}\hat{a}^{\dagger}_{c_{p(k)}}(t)\delta\hat{\Psi}_k(\bm{r},t).
\end{equation}
These operators scale as $\sqrt{N_{t_{p(k)}}(t)}$ [since $\hat{a}_{c_{p(k)}}(t)\sim\sqrt{N_{c_{p(k)}}(t)}$ and $\delta\hat{\Psi}_k(\bm{r},t)\sim\sqrt{N_{t_{p(k)}}(t)}$], which under our assumptions, are all small. 
This choice of fluctuation operator allows us to make an expansion of the Hamiltonian in terms of the number of condensate atoms, rather than the total number of atoms, i.e.\ we are not restricted to the assumption $N_{c}(t)\approx N$ [however, we must still satisfy $N_{c_{p{(k)}}}(t)\gg N_{t_{p{(k')}}}(t)$]. In addition, whilst the quasiparticle operators corresponding to $\Ltk$ are only approximately bosonic, from (\ref{equal0}) we see that the expectation value is exactly equal to zero \cite{gm}. These properties mark $\Ltk$ as an appropriate (although not perfect) expansion parameter. For a more in depth discussion on appropriate choices for fluctuation operators, we refer the reader to \cite{gm} and references therein. 

Through this choice of fluctuation operator, higher-order self-consistent equations of motion can be developed. This was precisely the case considered in \cite{gm} for a single component condensate, where the authors noted that the pair expectation values $\langle\Ltkdd\Ltkd\rangle$ always have a finite (and in general) non-zero value in an interacting gas. We call $\tilde{\Lambda}^{(\dagger)}$ (without a subscript) any member of the set of the (annihilation or creation) fluctuation operators, where it is not important which one it is.
An implication is that all equations of motion should be consistently taken to quadratic order in products of the fluctuation operators $\tilde{\Lambda}$ and $\tilde{\Lambda}^{\dagger}$.

For future use we note that the normal $\tilde{\Lambda}(\bm{r})$ pair is related to the normal $\delta\hat{\Psi}(\bm{r})$ pair (similarly to above we define $\delta\hat{\Psi}^{(\dagger)}$ to be any member of the set of all non-condensate field operators) by
\begin{equation}
	\label{lams}
	\Ltkdd\Ltkd=
	\frac{\left(\hat{a}^{\dagger}_{c_{p(k')}}(t)\hat{a}_{c_{p(k)}}(t)+\delta_{k,k'}^p\right)}{\sqrt{N_{c_{p(k)}}(t)N_{c_{p(k')}}(t)}}\delta\hat{\Psi}^{\dagger}_k(\bm{r}',t)\delta\hat{\Psi}_{k'}(\bm{r},t),
\end{equation}
and the exact commutation relation is given by
\begin{equation}
	\begin{split}
	\label{comm}
	[\Ltk,\Ltkddd]=&\frac{\hat{N}_{{c_{p(k)}}}(t)}{{N_{c_{p(k)}}}(t)}Q_{kk'}(\bm{r},\bm{r}',t)\\
	&-\frac{\delta_{k,k'}^p}{{N_{c_{p(k)}}}(t)}\delta\hat{\Psi}^{\dagger}_{k'}(\bm{r}',t)\delta\hat{\Psi}_k(\bm{r},t),
 \end{split}
\end{equation}
where we have been able to state that $\hat{N}_{{c_{p(k)}}}(t)=\hat{N}_{{c_{p(k')}}}(t)$ [and similarly ${N}_{{c_{p(k)}}}(t)={N}_{{c_{p(k')}}}(t)$] because of the presence of the Kronecker delta mapping term $\delta_{k,k'}^p$ that enforces components $k$ and $k'$ to be in the same set and hence having the same eigenvalue [see the definition of $Q_{kk'}(\bm{r},\bm{r}',t)$ in Eq. (\ref{qdeq})]. Note that in the second term of Eq. (\ref{comm}) we must still explicitly keep this Kronecker delta mapping term to account for the presence of the other terms $\delta\hat{\Psi}^{\dagger}_{k'}(\bm{r}',t)$ and $\delta\hat{\Psi}_{k}(\bm{r},t)$ which contain explicit $k$ and $k'$ index dependencies.

\subsection{Two- and three-component examples}
\label{form_6}

\subsubsection{Overview of key examples}
The above analysis has been kept entirely general. At this stage it is useful to summarise the analysis by means of a couple of specific examples, the first involving a two-component condensate and the second a three-component condensate. In the following we will make use of the short-hand notations: $\Psi_{jk}=\langle\hat{\Psi}^{\dagger}_{j}(\bm{r}')\hat{\Psi}_{k}(\bm{r})\rangle$,  $\phi_{jk}=\phi_j(\bm{r},t)\phi_k^*(\bm{r}',t)$, $\delta_{\bm{r}\bm{r}'}=\delta(\bm{r}-\bm{r}')$ and drop the arguments from the projectors, writing $Q_{jk}=Q_{jk}(\bm{r},\bm{r}',t)$. In Sec.\ \ref{egs} we will explore each of these examples in more detail.

\subsubsection{Two-component condensates}

For the two-component condensate we have $n=2$ and $C=2$, which corresponds to the cases (i) two mutually coherent components ($l=1$ with $|p_1|=2$) or (ii) two mutually incoherent components ($l=2$ with $|p_1|=1$ and $|p_2|=1$). 

{\it{Mutually Coherent Condensates: $l=1$}}. When both components are coherent we can write out the single-body density matrix, in the form of a 2 $\times$ 2 array of operator expectation values, as 
\begin{equation}
\varrho(\bm{r},\bm{r}',t) =
	\begin{pmatrix}
		\Psi_{11} & 
		\Psi_{21} \\
		\Psi_{12} & 
		\Psi_{22}		
	\end{pmatrix}.
\end{equation}
As both components are coherent, there is only one eigenvalue $N_{c_1}$, so that
\begin{equation}
	\int\begin{pmatrix}
		\Psi_{11} &
		\Psi_{21} \\
		\Psi_{12} & 
		\Psi_{22}		
	\end{pmatrix}
	\begin{pmatrix}
		\phi_{1}(\bm{r}',t)\\
		\phi_{2}(\bm{r}',t)
		\end{pmatrix}
		\dr'
		=N_{c_{{1}}}(t)
		\begin{pmatrix}
			\phi_1(\bm{r},t)\\
			\phi_2(\bm{r},t)
		\end{pmatrix}.
\end{equation}
We thus have a partitioning of the field operators defined by
\begin{equation}
	\begin{pmatrix}
		\hat{\Psi}_1(\bm{r})\\
		\hat{\Psi}_2(\bm{r})
	\end{pmatrix}
	=\hat{a}_{c_{1}}(t)
	\begin{pmatrix}
		\phi_1(\bm{r},t)\\
		\phi_2(\bm{r},t)
	\end{pmatrix}
	+
	\begin{pmatrix}
	\delta\hat{\Psi}_1(\bm{r},t)\\
	\delta\hat{\Psi}_2(\bm{r},t)
\end{pmatrix},
\end{equation}
where the creation operator is defined as
\begin{equation}
	\hat{a}^{\dagger}_{c_{1}}(t)=\int\left[\hat{\Psi}^{\dagger}_{1}(\bm{r})\phi_{1}(\bm{r},t)+\hat{\Psi}^{\dagger}_{2}(\bm{r})\phi_{2}(\bm{r},t)\right]\dr,
\end{equation}
and the noncondensate field operators are defined as
\begin{equation}
	\begin{pmatrix}
		\delta\hat{\Psi}_1(\bm{r},t)\\
		\delta\hat{\Psi}_2(\bm{r},t)
	\end{pmatrix}
	=\int
	\mathcal{Q}(\bm{r},\bm{r}',t)
	\begin{pmatrix}
		\hat{\Psi}_{1}(\bm{r}') \\
		\hat{\Psi}_{2}(\bm{r}')
	\end{pmatrix}
	\dr',
\end{equation}
with the matrix projector
\begin{equation}
	\mathcal{Q}(\bm{r},\bm{r}',t)=
	\begin{pmatrix}
		\delta_{\bm{rr}'}-\phi_{11} &
		-\phi_{12} \\
		-\phi_{21} &
		\delta_{\bm{rr}'}-\phi_{22}
	\end{pmatrix}.
\end{equation}
Finally we can quote the fluctuation operators:
\begin{equation}
	\begin{pmatrix}
		\tilde{\Lambda}_1(\bm{r},t)\\
		\tilde{\Lambda}_2(\bm{r},t)
	\end{pmatrix}
	=\frac{1}{\sqrt{N_{c_1}(t)}}\hat{a}^{\dagger}_{c_1}(t)
	\begin{pmatrix}
		\delta\hat{\Psi}_1(\bm{r},t)\\
		\delta\hat{\Psi}_2(\bm{r},t)
	\end{pmatrix}.
\end{equation}

{\it{Mutually Incoherent Condensates: $l=2$.}} In the incoherent case, we note that there are two subsets, $p_1$ and $p_2$, so that $\delta_{1,2}^p=0$. Our single-body density matrix then reads
\begin{equation}
	\varrho(\bm{r},\bm{r}',t)=
	\begin{pmatrix}
		\Psi_{11} & 0\\
		0 & \Psi_{22} 
	\end{pmatrix}.
\end{equation}
There are now two eigenvalues, $N_{c_1}$ and $N_{c_2}$ so that
\begin{equation}
	\int\begin{pmatrix}
		\Psi_{11} & 0\\
		0 & \Psi_{22} 
	\end{pmatrix}
	\begin{pmatrix}
		\phi_{1}(\bm{r}',t)\\
		\phi_{2}(\bm{r}',t)
		\end{pmatrix}
		\dr'=
		\begin{pmatrix}
			N_{c_{{1}}}(t)\phi_1(\bm{r},t)\\
			N_{c_{{2}}}(t)\phi_2(\bm{r},t)
		\end{pmatrix}.
\end{equation}
We thus have a partitioning of the field operators defined by
\begin{equation}
	\begin{pmatrix}
		\hat{\Psi}_1(\bm{r})\\
		\hat{\Psi}_2(\bm{r})
	\end{pmatrix}
	=
	\begin{pmatrix}
		\hat{a}_{c_{1}}(t)\phi_1(\bm{r},t)+\delta\hat{\Psi}_1(\bm{r},t)\\
		\hat{a}_{c_{2}}(t)\phi_2(\bm{r},t)+\delta\hat{\Psi}_2(\bm{r},t)
\end{pmatrix},
\end{equation}
where the creation operators are defined as
\begin{equation}
	\begin{pmatrix}
	\hat{a}^{\dagger}_{c_{1}}(t)\\
	\hat{a}^{\dagger}_{c_{2}}(t)
	\end{pmatrix}
	=\int
	\begin{pmatrix}
	\hat{\Psi}^{\dagger}_{1}(\bm{r})\phi_{1}(\bm{r},t)\\
	\hat{\Psi}^{\dagger}_{2}(\bm{r})\phi_{2}(\bm{r},t)
	\end{pmatrix}\dr,
\end{equation}
and the noncondensate field operators are defined as
\begin{equation}
	\begin{pmatrix}
		\delta\hat{\Psi}_1(\bm{r},t)\\
		\delta\hat{\Psi}_2(\bm{r},t)
	\end{pmatrix}
	=\int
\mathcal{Q}(\bm{r},\bm{r}',t)
	\begin{pmatrix}
		\hat{\Psi}_{1}(\bm{r}') \\
		\hat{\Psi}_{2}(\bm{r}')
	\end{pmatrix}
	\dr',
\end{equation}
with the (now diagonal) matrix projector
\begin{equation}
	\mathcal{Q}(\bm{r},\bm{r}',t)=
	\begin{pmatrix}
		\delta_{\bm{rr}'}-\phi_{11} &
		0 \\
		0 &
		\delta_{\bm{rr}'}-\phi_{22}
	\end{pmatrix}.
\end{equation}
Finally we can quote the fluctuation operators:
\begin{equation}
	\begin{pmatrix}
		\tilde{\Lambda}_1(\bm{r},t)\\
		\tilde{\Lambda}_2(\bm{r},t)
	\end{pmatrix}
	=
	\begin{pmatrix}
		\frac{1}{\sqrt{N_{c_1}(t)}}\hat{a}^{\dagger}_{c_1}(t)\delta\hat{\Psi}_1(\bm{r},t)\\
		\frac{1}{\sqrt{N_{c_2}(t)}}\hat{a}^{\dagger}_{c_2}(t)\delta\hat{\Psi}_2(\bm{r},t)
	\end{pmatrix}.
\end{equation}

\subsubsection{Three-component condensates}

Instead, if we have a three-component condensate then $n=3$ and $C=3$, which corresponds to the cases (i) three mutually coherent components ($l=1$ with $|p_1|=3$); (ii) two mutually coherent components and one incoherent component ($l=2$ with $|p_1|=2$ and $|p_2|=1$) or (iii) three mutually incoherent components ($l=3$ with $|p_1|=1$, $|p_2|=1$ and $|p_3|=1$). At this point we will only concentrate on case (ii) (the other two cases are similar to the two-component condensate cases above). We say that components $1$ and $2$ are mutually coherent (subset $p_1$) and component $3$ is incoherent with respect to the other two components (subset $p_2$), i.e.
\begin{equation}
	\mathsf{\hat{\Psi}}=\{\{\underbrace{\hat{\Psi}_1,\hat{\Psi}_2}_{p_1}\},\{\underbrace{\hat{\Psi}_3}_{p_2}\}\}.
\end{equation}	
	 Our single-body density matrix is then
\begin{equation}
	\varrho(\bm{r},\bm{r}',t)=
	\begin{pmatrix}
		\Psi_{11} &
		\Psi_{21} & 			0 \\
		\Psi_{12} &
		\Psi_{22} & 		0 \\
		0 &
		0 & \Psi_{33}
	\end{pmatrix}.
\end{equation}
There are two eigenvalues, $N_{c_1}$ associated with $p_1$ and $N_{c_2}$ associated with $p_2$, so that
\begin{equation}
	\int\begin{pmatrix}
		\Psi_{11} &
		\Psi_{21} & 			0 \\
		\Psi_{12} &
		\Psi_{22} & 		0 \\
		0 &
		0 & \Psi_{33}
	\end{pmatrix}
	\begin{pmatrix}
		\phi_{1}(\bm{r}',t)\\
		\phi_{2}(\bm{r}',t)\\
		\phi_{3}(\bm{r}',t)
		\end{pmatrix}
		\dr'=
		\begin{pmatrix}
			N_{c_{{1}}}(t)\phi_1(\bm{r},t)\\
			N_{c_{{1}}}(t)\phi_2(\bm{r},t)\\
			N_{c_{{2}}}(t)\phi_3(\bm{r},t)
		\end{pmatrix}.
\end{equation}
We thus have a partitioning of the field operators defined by
\begin{equation}
	\begin{pmatrix}
		\hat{\Psi}_1(\bm{r})\\
		\hat{\Psi}_2(\bm{r})\\
		\hat{\Psi}_3(\bm{r})
	\end{pmatrix}
	=
	\begin{pmatrix}
		\hat{a}_{c_{1}}(t)\phi_1(\bm{r},t)\\
		\hat{a}_{c_{1}}(t)\phi_2(\bm{r},t)\\
		\hat{a}_{c_{2}}(t)\phi_3(\bm{r},t)\\		
	\end{pmatrix}
	+
	\begin{pmatrix}
	\delta\hat{\Psi}_1(\bm{r},t)\\
	\delta\hat{\Psi}_2(\bm{r},t)\\
	\delta\hat{\Psi}_3(\bm{r},t)
\end{pmatrix},
\end{equation}
where the creation operators are defined as
\begin{equation}
	\begin{pmatrix}
	\hat{a}^{\dagger}_{c_{1}}(t)\\
	\hat{a}^{\dagger}_{c_{2}}(t)\\
	\end{pmatrix}
	=\int
	\begin{pmatrix}
	\hat{\Psi}^{\dagger}_{1}(\bm{r})\phi_{1}(\bm{r},t)+\hat{\Psi}^{\dagger}_{2}(\bm{r})\phi_{2}(\bm{r},t)\\
	\hat{\Psi}^{\dagger}_{3}(\bm{r})\phi_{3}(\bm{r},t)
	\end{pmatrix}\dr,
\end{equation}
and the noncondensate field operators are defined as
\begin{equation}
	\begin{pmatrix}
		\delta\hat{\Psi}_1(\bm{r},t)\\
		\delta\hat{\Psi}_2(\bm{r},t)\\
		\delta\hat{\Psi}_3(\bm{r},t)		
	\end{pmatrix}
	=\int
	\mathcal{Q}(\bm{r},\bm{r}',t)
	\begin{pmatrix}
		\hat{\Psi}_{1}(\bm{r}') \\
		\hat{\Psi}_{2}(\bm{r}')\\
		\hat{\Psi}_{3}(\bm{r}')
	\end{pmatrix}
	\dr',
\end{equation}
with the (block-diagonal) projector
\begin{equation}
	\mathcal{Q}(\bm{r},\bm{r}',t)=
	\begin{pmatrix}
		\delta_{\bm{rr}'}-\phi_{11} & -\phi_{12} &	0 \\
		-\phi_{21} &
		\delta_{\bm{rr}'}-\phi_{22} & 0 \\
		0 & 0 & \delta_{\bm{rr}'}-\phi_{33}
	\end{pmatrix}.
\end{equation}
 Finally we can quote the fluctuation operators:
\begin{equation}
	\begin{pmatrix}
		\tilde{\Lambda}_1(\bm{r},t)\\
		\tilde{\Lambda}_2(\bm{r},t)\\
		\tilde{\Lambda}_3(\bm{r},t)
	\end{pmatrix}
	=
	\begin{pmatrix}
		\frac{1}{\sqrt{N_{c_1}(t)}}\hat{a}^{\dagger}_{c_1}(t)\delta\hat{\Psi}_1(\bm{r},t)\\
		\frac{1}{\sqrt{N_{c_1}(t)}}\hat{a}^{\dagger}_{c_1}(t)\delta\hat{\Psi}_2(\bm{r},t)\\
		\frac{1}{\sqrt{N_{c_2}(t)}}\hat{a}^{\dagger}_{c_2}(t)\delta\hat{\Psi}_3(\bm{r},t)
	\end{pmatrix}.
\end{equation}

\section{Perturbative Expansion of the Effective Hamiltonian}
\label{pert}

\subsection{Overview of expansion}
\label{pert_1}

We wish to reformulate the effective Hamiltonian (\ref{Ham}) in terms of the fluctuation operators (\ref{fluc}). This will allow us to construct a perturbative expansion of the reformulated Hamiltonian in powers of the fluctuation operators (the small expansion parameters). This section will analyse in detail the form of the fluctuation operators and of the number operators, allowing us to make a consistent cubic approximation to the reformulated Hamiltonian. The cubic approximation will then be further developed in Sec.\ \ref{zero} and onwards. 

\subsection{Exact reformulation in terms of fluctuation operators}
\label{pert_2}

In the following we give the (exact) reformulation, according to (\ref{cond}) and (\ref{fluc}), of the Hamiltonian (\ref{Ham}), defining  $\tilde{U}_k=U_k\npk$ and $\Utjk=U_{jk}\sqrt{\npj\npk}$, and removing time arguments for brevity. We thus write
 $\hat{H}=\hat{H}^{(\tilde{\Lambda}^0)}+\hat{H}^{(\tilde{\Lambda}^1)}+\hat{H}^{(\tilde{\Lambda}^2)}+\hat{H}^{(\tilde{\Lambda}^3)}+\hat{H}^{(\tilde{\Lambda}^4)}$, where the expressions for $\hat{H}^{(\tilde{\Lambda}^i)}$, and brief calculational details to obtain each term, are given in Appendix \ref{reform}. The terms of this reformulated Hamiltonian are arranged in orders of powers of products of the fluctuation operators, so that the term $\hat{H}^{(\tilde{\Lambda}^i)}$ contains products of $\tilde{\Lambda}$ of order $i$ (where $i\le4$). Our aim is to reduce this Hamiltonian so that we obtain a lowest-order consistent dynamical representation of the $n$-component condensate. We will see later that this implies a reduction of the above Hamiltonian to a third-order Hamiltonian (as in the case of a single component \cite{gm}), but before we do, we first need to find an approximation to the number and fluctuation operators.

\subsection{Approximations to number and fluctuation operators}
\label{pert_3}

The reformulated Hamiltonian (\ref{Hamil}) can be reduced to a third-order Hamiltonian by means, firstly, of the expansion of the condensate number operators and, secondly, by means of a Gaussian approximation to the fluctuation terms. Each subset of $\mathsf{\hat{\Psi}}$ is in a number eigenstate, having particle number $\npk$ [with associated mapping $p(k)$]. This implies that the number fluctuations of the condensate and noncondensate components within each subset must be equal and opposite (see Appendix \ref{app_1}). To zeroth- (and first-) order in the fluctuation operators, $\nhpk=\npk$, whereas to second-order we have
\begin{equation}
	\label{numapp}
	\nhpk=\npk+\sum_{k'=1}^{n}\delta_{k,k'}^p\int \biggl[ \left<\tilde{\Lambda}^{\dagger}_{k'}(\bm{r})\tilde{\Lambda}_{k'}(\bm{r})\right>
	-\tilde{\Lambda}^{\dagger}_{k'}(\bm{r})\tilde{\Lambda}_{k'}(\bm{r})\biggr] \dr.
\end{equation}

We can now use (\ref{numapp}) to express the commutation relation $[\tilde{\Lambda}_k(\bm{r}), \tilde{\Lambda}_{k'}^{\dagger}(\bm{r}')]$ (\ref{comm}) in terms of the condensate numbers and expectation values of $\tilde{\Lambda}_k(\bm{r})$ and $\tilde{\Lambda}_{k'}^{\dagger}(\bm{r}')$: to a Gaussian level of approximation we may replace pairwise products of the fluctuation operators $\tilde{\Lambda}(\bm{r})$ and $\tilde{\Lambda}^{\dagger}(\bm{r})$ by their expectation values \cite{gm}. We thus write (see Appendix \ref{app_1})
\begin{equation}
	\begin{split}
[\Lk,\Lkddd]&\approx Q_{kk'}(\bm{r},\bm{r}')-\frac{\left<\Lkddd\Lk\right>}{\npk}\delta_{k,k'}^p,
\end{split}
\label{flucapp}
\end{equation}
whereas to zeroth and first order, the commutator may be approximated by 
\begin{equation}
[\Lk,\Lkddd]\approx Q_{kk'}(\bm{r},\bm{r}').
\label{fluccomzero}
\end{equation}

\subsection{Reduction to a third-order Hamiltonian}
\label{pert_4}

The final step in constructing a third-order Hamiltonian from the full Hamiltonian (\ref{Hamil}) is to consistently deal with the cubic and quartic powers of products of the fluctuations operators. It turns out that the quartic terms can be safely neglected (see an equivalent discussion in \cite{gm}), whereas the cubic terms require a Hartree--Fock factorisation. In general, the Hartree--Fock factorisation is written as \cite{gm,hutch}
\begin{equation}
	\begin{split}
	\label{htf}
	\tilde{\Lambda}_k^{\dagger}(\bm{r})\tilde{\Lambda}_{k'}(\bm{r}')\tilde{\Lambda}_{k''}(\bm{r}'')\approx&\left<\tilde{\Lambda}_k^{\dagger}(\bm{r})\tilde{\Lambda}_{k'}(\bm{r}')\right>\tilde{\Lambda}_{k''}(\bm{r}'')\\
	&+\left<\tilde{\Lambda}_k^{\dagger}(\bm{r})\tilde{\Lambda}_{k''}(\bm{r}'')\right>\tilde{\Lambda}_{k'}(\bm{r}')\\
	&+\left<\tilde{\Lambda}_{k'}(\bm{r}')\tilde{\Lambda}_{k''}(\bm{r}'')\right>\tilde{\Lambda}_k^{\dagger}(\bm{r}),
\end{split}
\end{equation}
i.e.\ the cubic products are re-expressed as linear terms modified by a pair average.

We can now substitute, consistently, (\ref{numapp}), (\ref{flucapp}) and (\ref{htf}) into (\ref{Hamil}) to give the third-order Hamiltonian $\hat{H}_3$.
For convenience of notation, we split this Hamiltonian into a sum of terms in orders of the powers of the number eigenvalues (again we denote the set of number eigenvalues as $N_c$). So we write $\hat{H}_3=\hat{H}_3^{(N_c^1)}+\hat{H}_3^{(N_c^{1/2})}+\hat{H}_3^{(N_c^0)}+\hat{H}_3^{(N_c^{-1/2})}$ and further split $\hat{H}_3^{(N_c^0)}=\hat{H}_3^{(N_c^0)_{a}}+\hat{H}_3^{(N_c^0)_{b}}$ and $\hat{H}_3^{(N_c^{-1/2})}=\hat{H}_3^{(N_c^{-1/2})_a}+\hat{H}_3^{(N_c^{-1/2})_b}$. These expressions, and further details of the calculation, are given in Appendix \ref{reform_cubic}. Note that $\hat{H}=\hat{H}_3+\mathcal{O}(\tilde{\Lambda}^4(\bm{r}),N_{c}^{-1})$. 

\section{Evolution equations}
\label{evolu}

\subsection{Overview of approach}
\label{evolu_1}

The rather unwieldy cubic Hamiltonian (\ref{Ham3_1}) forms the basis of our approach to obtain a consistent set of dynamical equations for the mutually coherent and incoherent condensate and non-condensate parts. To develop these, we will first require expressions for the time evolution of the fluctuation and number operators. We derive these in the next two subsections.

\subsection{Evolution of the fluctuation operators}
\label{evolu_2}

In general the Heisenberg time evolution of the fluctuation operators is given by
\begin{equation}
	\label{lambda_time}
	\rmi\hbar\frac{d}{dt}\Lk=[\Lk,\hat{H}]+\rmi\hbar\frac{\partial}{\partial t}\Lk,
\end{equation}
where (see Appendix \ref{app_2}),
\begin{widetext}
\begin{equation}
	\label{fluc_evol}
	\begin{split}
			\rmi\hbar\frac{\partial}{\partial t}\Lk=&\skd1n\Bigg(-\sqrt{\npk}\int Q_{kk'}(\bm{r},\bm{r}')\left[\rmi\hbar\frac{\partial\phi_{k'}(\bm{r}')}{\partial t}\right]\dr'
			+\delta_{k,k'}^p\Bigg\{-\phi_k(\bm{r})\int\left[\rmi\hbar\frac{\partial\phi_{k'}^*(\bm{r}')}{\partial t}\right]\tilde{\Lambda}_{k'}(\bm{r}')\dr'\\
			&+\tilde{\Lambda}_k(\bm{r})\int\left[\rmi\hbar\frac{\partial\phi_{k'}(\bm{r}')}{\partial t}\right]\phi_{k'}^*(\bm{r}')\dr'
			+\frac{1}{\sqrt{\npk}}\int\left[\rmi\hbar\frac{\partial\phi_{k'}(\bm{r}')}{\partial t}\right]
			\left<\tilde{\Lambda}^{\dagger}_{k'}(\bm{r}')\tilde{\Lambda}_{k}(\bm{r})\right>\dr'\Bigg\}\Bigg),
	\end{split}
\end{equation}
\end{widetext}
and where we have used (\ref{numapp}) to approximate the number operators in terms of expectation values of the fluctuation operators. Equation (\ref{fluc_evol}) gives us the explicit time evolution of the fluctuation operators.

\subsection{Evolution of the number operators}
\label{evolu_3}

Similarly, we require the time evolution of the number operators, which can be found from 
\begin{equation}
	\label{num_time}
	\rmi\hbar\frac{d}{dt}\nhpk=[\nhpk,\hat{H}]+\rmi\hbar\frac{\partial}{\partial t}\nhpk,
\end{equation}
where $\hat{N}_{c_{p(k)}}(t)\equiv\hat{a}^{\dagger}_{c_{p(k)}}(t)\hat{a}_{c_{p(k)}}(t)$. Following a similar procedure to above, invoking (\ref{a_delta}), we obtain
\begin{equation}
	\label{num_evol}
	\begin{split}
	\rmi\hbar\frac{\partial}{\partial t}\nhpk=&\sqrt{\npk}\sum_{k'=1}^{n}\delta_{k,k'}^p\bigg\{\int\tilde{\Lambda}_{k'}^{\dagger}(\bm{r})\left[\rmi\hbar\frac{\partial\phi_{k'}(\bm{r})}{\partial t}\right]\\
	&+\left[\rmi\hbar\frac{\partial\phi_{k'}^*(\bm{r})}{\partial t}\right]\tilde{\Lambda}_{k'}(\bm{r})\dr\bigg\}.
\end{split}
\end{equation}
We thus see that $\langle\rmi\hbar \partial \nhpk/ \partial t\rangle=0$, and so
\begin{equation}
\rmi\hbar\frac{d}{dt}\npk=\left<\rmi\hbar\frac{d}{dt}\nhpk\right>
=\left<[\nhpk,\hat{H}]\right>
\end{equation}
is our equation for the time evolution of the component number. 

\section{Zeroth-, first- and second-order approximations to the Hamiltonian}
\label{zero}

\subsection{Overview of resulting dynamical equations}
Following the derivation of the evolution equations for both the fluctuation operators (\ref{fluc_evol}) and the condensate numbers (\ref{num_evol}), this section will analyse  the Hamiltonian (\ref{Ham3_1}) up to a second-order approximation. This provides, in the first-order approximation, a derivation for the time-dependent Gross--Pitaevskii equations as the sole equations governing the multi-component condensate dynamics (so not capturing any out-of-condensate dynamics). The the second-order approximation provides a derivation for equations governing the out of condensate dynamics --- called the modified Bogoliubov--de Gennes equation (together with its complex conjugate), in multi-component form, which are coupled to the multicomponent Gross--Pitaevskii equations. The condensate and non-condensate dynamics yielded by this system of equations are, however,  not self-consistent with regard to the particle numbers.  We will therefore require a third-order (i.e.\ full) treatment of the Hamiltonian (\ref{Ham3_1}). The third-order treatment will be considered in Sec.\ \ref{there}.

\subsection{Zeroth order}
\label{zero_1}

The lowest order approximation to the cubic Hamiltonian (\ref{Ham3_1}) is to neglect any terms involving the fluctuation operators. In the zeroth order we then take $\hat{H}_3$ and neglect any fluctuation terms. This gives the zeroth-order Hamiltonian, 
which is identical to $\hat{H}_3^{(N_c^1)}$, but here denoted $H_0$,  where
\begin{equation}
	\label{Ham0}
	\begin{split}
	H_0=&\int\sk1n\Bigg\{\npk\phi_k^*(\bm{r})\left[H_{\text{sp}}^k(\bm{r})+\frac{\Utkk}{2}|\phi_k(\bm{r})|^2\right]\phi_k(\bm{r})
	\\&
	+\sjk\Utjk\sqrt{\npj\npk}|\phi_j(\bm{r})|^2|\phi_k(\bm{r})|^2
	\\&
	+\sum_{\substack{j=1\\j\neq k}}^n\npk\phi_j^*(\bm{r})H_{\text{ob}}^{jk}(\bm{r})\phi_k(\bm{r})\Bigg\}\dr,
	\end{split}
\end{equation}
and is purely classical (hence we write $\hat{H}_0\equiv H_0$). In terms of a mean-field theory in which the $n$-component condensate is assumed to be at absolute zero and with all bosons condensed (i.e.\ strictly zero noncondensate bosons), the appropriate Hamiltonian [noting that it is common to scale the condensate mode(s) to be normalised to the number of condensate particles] is given by this zeroth-order Hamiltonian \cite{tsubota,ma}. Noncondensate particles are not accounted for in this Hamiltonian and our analysis by consequence then proceeds to higher-order approximations.
 
\subsection{First order}
\label{zero_2}

To the next order of approximation, a first-order approximation on $\hat{H}_3$, we consider terms up to linear order in the fluctuation operators. At this level of approximation the appropriate Hamiltonain is given by $\hat{H}_1=\hat{H}_3^{(N_c^1)}+\hat{H}_3^{(N_c^{1/2})}$. Working with this Hamiltonian and Eq. (\ref{lambda_time}) we will obtain Gross--Pitaevskii equations describing the evolution of the condensate modes. Details of the calculations are given in Appendix \ref{app_1st} and we note that the set of time-dependent Gross--Pitaevskii (GP) equations read
\begin{subequations}
	\label{order1}
\begin{multline}
	\label{gp_1}
		\rmi\hbar\frac{\partial\phi_k(\bm{r})}{\partial t}=\Bigg[H_{\text{sp}}^k(\bm{r})+\Utkk|\phi_k(\bm{r})|^2
		+\sjdkdn\sqrt{\frac{\npj}{\npk}}\Utjk|\phi_{j}(\bm{r})|^2
		\\ -\lambda_0^{p(k)}\Bigg]\phi_k(\bm{r})+\sjdkdn H_{\text{ob}}^{kj}(\bm{r})\phi_j(\bm{r}),
\end{multline}
where
\begin{equation}
	\label{cp_1}
	\begin{split}
		\lambda_0^{p(k)}&=\int\sum_{k'=1}^n\Bigg\{\delta_{k,k'}^p\phi_{k'}^*(\bm{r})\Bigg[H_{\text{sp}}^{k'}(\bm{r})+\tilde{U}_{k'k'}|\phi_{k'}(\bm{r})|^2\\
		&\quad-\rmi\hbar\frac{\partial}{\partial t}+\sum_{\substack{j=1\\j\neq k'}}^n\tilde{U}_{jk'}\sqrt{\frac{N_{c_{p(j)}}}{N_{c_{p(k')}}}}|\phi_{j}(\bm{r})|^2\Bigg]\\
		&\quad+\sum_{\substack{j=1\\j\neq k'}}^n\delta_{j,k}^p\phi_j^*(\bm{r})H_{\text{ob}}^{jk'}(\bm{r})\Bigg\}\phi_{k'}(\bm{r})\dr.
	\end{split}
\end{equation}
\end{subequations}

System (\ref{order1}) is the Gross--Pitaevskii equation for condensate part $\phi_k$, with an associated nonlinear eigenvalue $\lambda_0^{p(k)}$ that has the appearance of a chemical potential. Note that an identical set of equations would result from the zeroth-order Hamiltonian $H_0$ (\ref{Ham0}). There are then $n$ Gross--Pitaevskii equations (one for each condensate part) and the number of distinct nonlinear eigenvalues depends on the number $l$ of subsets of $\mathsf{\hat{\Psi}}$. Note that these $\lambda_0^{p(k)}$ are real.

However, whilst this system of equations is often used as the lowest order representation of the multi-component condensate \cite{tsubota,ma}, it does not capture any of the non-condensate dynamics (in fact one can show, see \cite{gm} for details, that to this order there is no time-dependence on the non-condensate components). We are thus required to go to higher-order approximations.

\subsection{Second order}
\label{zero_3}

As a second-order approximation to the cubic Hamiltonian $\hat{H}_3$, we keep terms up to and including those quadratic in the fluctuation operators. 
In the same vein to the first-order calculations above, we can use the second-order approximation to (\ref{fluc_evol}), which means we should calculate
\begin{multline}
	\label{sub}
	\rmi\hbar\frac{d\Lk}{dt}=[\Lk,\hat{H}_2]+\skd1n\Bigg(-\sqrt{\npk}\\
\begin{split}
	&\times\int Q_{kk'}(\bm{r},\bm{r}')\left[\rmi\hbar\frac{\partial\phi_{k'}(\bm{r}')}{\partial t}\right]\dr'\\
			&+\delta_{k,k'}^p\Bigg\{-\phi_k(\bm{r})\int\left[\rmi\hbar\frac{\partial\phi_{k'}^*(\bm{r}')}{\partial t}\right]\tilde{\Lambda}_{k'}(\bm{r}')\dr'\\
			&+\tilde{\Lambda}_k(\bm{r})\int\left[\rmi\hbar\frac{\partial\phi_{k'}(\bm{r}')}{\partial t}\right]\phi_{k'}^*(\bm{r}')\dr'\Bigg\}\Bigg).
\end{split}
\end{multline}	
As before it is relatively straightforward to obtain the expression for the commutator $[\Lk,\hat{H}_2]$: using (\ref{fluccomzero}) the terms linear and quadratic in the fluctuation operators can be dealt with as previously, while the terms cubic in the fluctuation operator reduce to linear form. We do not provide the explicit expression for $d\Lk/dt$, instead choosing to skip to the expression that results after having taken the expectation value of (\ref{sub}).  We are then left with the same GP equations of (\ref{gp_1}) and associated nonlinear eigenvalues (\ref{cp_1}). We can use this fact to substitute the GP equations (\ref{gp_1}) into the right-hand side of (\ref{sub}). Doing this, we get
\begin{multline}
	\label{mbdg}
	\rmi\hbar\frac{d\Lk}{dt}=\Bigg[H_{\text{sp}}^k(\bm{r})+\tilde{U}_{kk}|\phi_k(\bm{r})|^2+\sjdkdn\tilde{U}_{jk}\sqrt{\frac{\npj}{\npk}}|\phi_{j}(\bm{r})|^2\\
	\begin{split}
	&-\lambda_0^{p(k)}\Bigg]\Lk+\sjdkdn H_{\text{ob}}^{kj}(\bm{r})\Lj\\
	&+\int\skd1n Q_{kk'}(\bm{r},\bm{r}')\Bigg\{\Utkdkd\left[\phi_{k'}^*(\bm{r}')\tilde{\Lambda}_{k'}(\bm{r}')+\text{H.c.}\right]\\
	&+\sum_{\substack{j=1\\j\neq k'}}^n\tilde{U}_{jk'}\left[\phi_j^*(\bm{r}')\tilde{\Lambda}_{j}(\bm{r}')+\text{H.c.}\right]\Bigg\}\phi_{k'}(\bm{r}')\dr'.
	\end{split}
\end{multline}
This and its Hermitian conjugate form the modified Bogoliubov--de Gennes (MBdG) equations \cite{pierre}. With this, we can calculate the evolution of the noncondensate parts: we note that, from (\ref{n_app}), $N_{c_{p(k)}}=N-\int\sum_{k'=1}^{n}\langle\tilde{\Lambda}^{\dagger}_{k'}(\bm{r})\tilde{\Lambda}_{k'}(\bm{r})\rangle\delta_{k,k'}^p\dr$, so that
\begin{equation}
	\label{cond_mod}
	\begin{split}	\rmi\hbar\frac{d\npk}{dt}=&\int\sum_{k'=1}^{n}
		\delta_{k,k'}^p\Bigg\{\tilde{U}_{k'k'}\left[\phi_{k'}^{*^2}(\bm{r})\left<\tilde{\Lambda}_{k'}^2(\bm{r})\right>-\text{H.c.}\right]\\
	&+\sum_{\substack{j=1\\j\neq k'}}^n\tilde{U}_{jk'}\bigg[\phi_{j}^*(\bm{r})\phi_{k'}^*(\bm{r})\left<\Lj\Lkd\right>-\text{H.c.}\bigg]\Bigg\}\dr,
\end{split}
\end{equation}
where we have used the modified Bogoliubov--de Gennes equations from above.

Our second-order system thus comprises the Gross--Pitaevskii equations (\ref{gp_1}) with associated nonlinear eigenvalues (\ref{cp_1}) together with the modified Bogoliubov--de Gennes equations (\ref{mbdg}). There are two key issues, however, that one must highlight before proceeding. The first is the appearance of terms quadratic in the fluctuation operators in the evolution of the condensate number of (\ref{cond_mod}), which we have, up to now, consistently neglected when dealing with the second-order Hamiltonian. This ``inconsistency'' [if we are to retain (\ref{cond_mod})] leads us to the second key issue, which is the possibility, in an out-of-equilibrium evolution, for unconstrained growth of the non-condensate part without there being any corresponding effect on the condensate evolution. This unconstrained growth is a result of the one-way condensate and noncondensate part dynamics involved in the coupled system of the Gross--Pitaevskii equations (\ref{gp_1}) and the modified Bogoliubov--de Gennes equations (\ref{mbdg}): the Gross--Pitaevskii equations (derived through a first-order treatment of the effective Hamiltonian), which evolve the condensate parts, explicitly contain only condensate part terms, whereas the modified Bogoliubov--de Gennes equations (derived through a second-order treatment of the effective Hamiltonian), which evolve the noncondensate parts, contain both condensate and noncondensate parts. This allows for an unconstrained growth in the noncondensate part terms through evolution of the modified Bogoliubov--de Gennes equations without any corresponding effect on the condensate parts in the Gross--Pitaevskii equations (see \cite{bmg,gm} for more details concerning unconstrained growth in a single component condensate). We are thus left with a non-self-consistent set of dynamical equations. To what extent this system can be considered appropriate depends very much on the actual dynamical configuration. Irrespective, a treatment of the cubic Hamiltonian (\ref{Ham3_1}) will enable us to form a self-consistent set of dynamical equations for the condensate and non-condensate parts.

\section{Second-order equations of motion}
\label{there}

In light of the inconsistent nature of the first-order approximation to the dynamical equations resulting from a second-order Hamiltonian, we proceed with a third-order approximation to the Hamiltonian. This is the highest order that we will be required to go to in order to achieve a self-consistent set of dynamical equations for the condensate and noncondensate parts: we will find that the equation governing the condensate parts [the Gross--Pitaevskii equation in a first- and second-order treatment, (\ref{gp_1}), with only condensate part dependence] is, in this third-order treatment, generalised to be dependent on both the condensate and noncondensate parts. This higher-order Gross--Pitaevskii equation we refer to as the generalised Gross--Pitaevskii (GGP) equation. It will be shown to be coupled, in a self-consistent manner, to the modified Bogoliubov--de Gennes equations of (\ref{mbdg}), which in the third-order treatment remain unchanged.

To a third-order approximation, the effective Hamiltonian has already been written down, and referred to as the cubic Hamiltonian $\hat{H}_3$, in (\ref{Ham3_1}). In dealing with the cubic Hamiltonian, we must use the full form of (\ref{fluc_evol}), which means that we need to consider
\begin{equation}
	\begin{split}
	\rmi\hbar\frac{d\Lk}{dt}=&[\Lk,\hat{H}_3]+\skd1n\Bigg(-\sqrt{\npk}\\
	&\times\int Q_{kk'}(\bm{r},\bm{r}')\left[\rmi\hbar\frac{\partial\phi_{k'}(\bm{r}')}{\partial t}\right]\dr'\\
			&+\delta_{k,k'}^p\Bigg\{-\phi_k(\bm{r})\int\left[\rmi\hbar\frac{\partial\phi_{k'}^*(\bm{r}')}{\partial t}\right]\\
			&\times\tilde{\Lambda}_{k'}(\bm{r}')\dr'+\tilde{\Lambda}_k(\bm{r})\int\left[\rmi\hbar\frac{\partial\phi_{k'}(\bm{r}')}{\partial t}\right]\\
			&\times\phi_{k'}^*(\bm{r}')\dr'+\frac{1}{\sqrt{\npk}}\int\left[\rmi\hbar\frac{\partial\phi_{k'}(\bm{r}')}{\partial t}\right]\\
			&\times\left<\tilde{\Lambda}^{\dagger}_{k'}(\bm{r}')\tilde{\Lambda}_{k}(\bm{r})\right>\dr'\Bigg\}\Bigg).
	\label{H3}
\end{split}
\end{equation}
As explicitly written down in the cubic Hamiltonian (\ref{Ham3_1}) and in the above equation, we now consider terms of order $\mathcal{O}(N_c^{-1/2})$.
Calculation of $[\Lk,\hat{H}_3]$ is only slightly more involved than before: the only additional terms that result from use of the cubic, rather than second-order Hamiltonian, are all of order $\mathcal{O}(\tilde{\Lambda}/N_c^{1/2})$, i.e.\ they are linear in the fluctuation operators which makes calculation of the commutator rather straightforward. The analysis on (\ref{H3}) becomes fairly cumbersome, although straightforward, and so we do not quote it in its entirety in the main text. Instead we will skip to the resultant expression: this is obtained by taking the expectation value of (\ref{H3}) and inserting the Gross--Pitaevskii equation (\ref{gp_1}) on the fourth and fifth lines of (\ref{H3}). After some calculation we arrive eventually to an equation for the evolution of the condensate parts, named the generalised Gross--Pitaevskii (GGP) equation:
\begin{widetext}
\begin{subequations}
\begin{equation}
	\label{ggp}
	\begin{split}
		\rmi\hbar\frac{\partial\phi_k(\bm{r})}{\partial t}=&\Bigg\{H_{\text{sp}}^k(\bm{r})+\tilde{U}_{kk}\left[\left(1-\frac{1}{\npk}\right)|\phi_k(\bm{r})|^2+2\frac{\left<\Lkdag\Lk\right>}{\npk}\right]+\sum_{\substack{j=1\\j\neq k}}^n\tilde{U}_{jk}\bigg[\sqrt{\frac{N_{c_{p(j)}}}{\npk}}\left(1-\frac{\delta_{j,k}^p}{\npk}\right)|\phi_{j}(\bm{r})|^2\\
		&+
	\frac{\left<\Ljdag\Lj\right>}{\sqrt{\npj\npk}}\bigg]-\lambda_2^{p(k)}\Bigg\}\phi_k(\bm{r})+\frac{\Utkk}{\npk}\left<\tilde{\Lambda}_k^2(\bm{r})\right>\phi_k^*(\bm{r})+\sum_{\substack{j=1\\j\neq k}}^n\bigg\{\frac{\Utjk}{\npk}\bigg[\left<\Ljdag\Lk\right>\phi_j(\bm{r})\\
	&+\left<\Lj\Lk\right>\phi_j^*(\bm{r})\bigg]+H_{\text{ob}}^{kj}(\bm{r})\phi_j(\bm{r})\bigg\}-\int\skd1n\delta_{k,k'}^p\frac{|\phi_{k'}(\bm{r}')|^2}{\npkd}\Bigg\{\tilde{U}_{k'k'}\bigg[\left<\tilde{\Lambda}_{k'}(\bm{r}')\tilde{\Lambda}_k(\bm{r})\right>\phi_{k'}^*(\bm{r}')\\
	&+\left<\tilde{\Lambda}_{k'}^{\dagger}(\bm{r}')\tilde{\Lambda}_k(\bm{r})\right>\phi_{k'}(\bm{r}')\bigg]+\sum_{\substack{j=1\\j\neq k'}}^n \tilde{U}_{jk'}\left[\left<\tilde{\Lambda}_{j}(\bm{r}')\tilde{\Lambda}_k(\bm{r})\right>\phi_{j}^*(\bm{r}')+\left<\tilde{\Lambda}_{j}^{\dagger}(\bm{r}')\tilde{\Lambda}_k(\bm{r})\right>\phi_{j}(\bm{r}')\right]\Bigg\}\dr',
	\end{split}
\end{equation}
where $\lambda_2^{p(k)}$ is a nonlinear eigenvalue given by
\begin{equation}
	\label{lam222}
	\begin{split}
	\lambda_2^{p(k)}=&\int\skd1n\Bigg(\delta_{k,k'}^p\phi_{k'}^*(\bm{r})\Bigg\{H_{\text{sp}}^{k'}(\bm{r})+\tilde{U}_{k'k'}\left[\left(1-\frac{1}{\npkd}\right)|\phi_{k'}(\bm{r})|^2+\frac{2}{\npkd}\left<\tilde{\Lambda}_{k'}^{\dagger}(\bm{r})\tilde{\Lambda}_{k'}(\bm{r})\right>\right]-\rmi\hbar\frac{\partial}{\partial t}\\
	&+\sjkdn\tilde{U}_{jk'}\left\{\sqrt{\frac{\npj}{\npkd}}\left[1-\frac{\delta_{j,k}^p}{\npkd}\right]|\phi_j(\bm{r})|^2+\frac{\left<\Ljdag\Lj\right>}{\sqrt{\npj\npkd}}\right\}\Bigg\}\phi_{k'}(\bm{r})+\sum_{\substack{j=1\\j\neq k'}}^n \delta_{j,k}^p\phi_{j}^*(\bm{r}) H_{\text{ob}}^{jk'}(\bm{r})\phi_{k'}(\bm{r})\\
	&+\delta_{k,k'}^p\frac{\tilde{U}_{k'k'}}{\npkd}\left<\tilde{\Lambda}_{k'}^2(\bm{r})\right>\phi_{k'}^{*^2}(\bm{r})+\sjkdn\delta_{k,k'}^p\frac{\tilde{U}_{jk'}}{{\npkd}}\left[\phi_j^*(\bm{r})\left<\Lj\Lkd\right>+\phi_j(\bm{r})\left<\Ljdag\Lkd\right>\right]\phi_{k'}^*(\bm{r})\Bigg)\dr.
	\end{split}
\end{equation}
\end{subequations}
\end{widetext}
Equations (\ref{ggp}) and (\ref{lam222}) constitute our generalised Gross--Pitaevskii equation. This should be contrasted with the lower order Gross--Pitaevskii equation of (\ref{gp_1}). The above expression for $\lambda_2^{p(k)}$ is also to be contrasted with the expression for the (real) nonlinear eigenvalues $\lambda_0^{p(k)}$ (\ref{cp_1}). 
The following calculation shows that they have a non-zero imaginary part:
\begin{equation}
	\begin{split}
	\lambda_2^{p(k)}-\left(\lambda_2^{{p(k)}}\right)^*=&\frac{1}{\npk}\int\sum_{k'=1}^{n}
		\delta_{k,k'}^p\Bigg\{\tilde{U}_{k'k'}\bigg[\phi_{k'}^{*^2}(\bm{r})\\
		&\times\left<\tilde{\Lambda}_{k'}^2(\bm{r})\right>-\text{H.c.}\bigg]+\sum_{\substack{j=1\\j\neq k'}}^n\tilde{U}_{jk'}\bigg[\phi_{j}^*(\bm{r})\\
		&\times\phi_{k'}^*(\bm{r})\left<\Lj\Lkd\right>-\text{H.c.}\bigg]\Bigg\}\dr,
		\end{split}
\end{equation}
which gives that, using (\ref{cond_mod}),
\begin{equation}
	\frac{d\npk}{dt}=\frac{1}{\rmi\hbar}\left[\lambda_2^{p(k)}-\left(\lambda_2^{{p(k)}}\right)^*\right]\npk.
\end{equation}

We can now state the final form of our self-consistent dynamical set of equations for the condensate and non-condensate parts. The condensate parts (one for each component) are governed by the time-dependent generalised Gross--Pitaevskii equations (\ref{ggp}) with associated nonlinear eigenvalues given in (\ref{lam222}). A higher (third-) order treatment of the Hamiltonian to obtain the noncondensate part evolution shows that the form of the evolution is the same as that given by the modified Bogoliubov--de Gennes equations of (\ref{mbdg}). Thus, the generalised Gross--Pitaevskii equations are coupled in a consistent way to the modified Bogoliubov--de Gennes equations (\ref{mbdg}) with associated nonlinear eigenvalues (\ref{cp_1}). In the next section we will give specific examples of use of these expressions for $n=1$, $n=2$ and $n=3$.

\section{Examples}
\label{egs}

\subsection{Overview of chosen examples}
It is instructive to present a few example systems, the first a trivial reduction to a single component, then two-component systems, and finally a three-component system. In each case we will consider possible subsets of $\mathsf{\hat{\Psi}}$, for which we will write down the (i) coupled GP equations and (ii) the coupled GGP equations and the MBdG equations (these examples correspond to the examples given in the discussion of the density operator in Sec.\ \ref{form_6}).

\subsection{Single-component condensates}

The number-conserving approach to a single component condensate has been considered in detail in \cite{gm}. We show that the formalism presented for the $n$-component condensate in this paper recovers the single component condensate system of \cite{gm}. In the case where there is only a single component, $n=1$, we define $p(1)\equiv 1$, and the lowest order approximation to the equations of motion result in the GP equation (\ref{order1}):
\begin{subequations}
	\label{order1_ex_1_1}
\begin{equation}
	\label{gp_ex_1_1}
		\rmi\hbar\frac{\partial\phi_1(\bm{r})}{\partial t}=\left[H_{\text{sp}}^1(\bm{r})+\tilde{U}_{11}|\phi_1(\bm{r})|^2-\lambda_0^1\right]\phi_1(\bm{r}),
\end{equation}
with nonlinear eigenvalue
\begin{equation}
	\label{cp_1_ex_1_1}
	\lambda_0^1=\int\phi_{1}^*(\bm{r})\left[H_{\text{sp}}^{1}(\bm{r})+\tilde{U}_{11}|\phi_{1}(\bm{r})|^2-\rmi\hbar\frac{\partial}{\partial t}\right]\phi_{1}(\bm{r})\dr.
\end{equation}
\end{subequations}
This GP equation and associated nonlinear eigenvalue are identical to (57) and (58) of \cite{gm}. Note that this system is the commonly used GP equation applied to a single component condensate when in the zero-temperature limit \cite{ps}.

To the next order, the resulting dynamical equations of motion are the MBdG equations (\ref{mbdg})
\begin{multline}
	\label{mbdg_ex_1_1}
	\rmi\hbar\frac{d\tilde{\Lambda}_1(\bm{r})}{dt}=\left[H_{\text{sp}}^1(\bm{r})+\tilde{U}_{11}|\phi_1(\bm{r})|^2-\lambda_0^1\right]\tilde{\Lambda}_1(\bm{r})\\
		+\int Q_{11}(\bm{r},\bm{r}')\tilde{U}_{11}\bigg[\phi_{1}^*(\bm{r}')\tilde{\Lambda}_1(\bm{r}')+\text{H.c.}\bigg]\phi_{1}(\bm{r}')\dr',
\end{multline}
coupled to the above GP equation (\ref{order1_ex_1_1}).
Finally, the second order (consistent) dynamical equations of motion couple this MBdG equation (\ref{mbdg_ex_1_1}) to the GGP equation (\ref{ggp}), written for $n=1$ as
\begin{subequations}
	\label{order23_ex_1_1b}
\begin{equation}
	\label{ggp_ex_1}
	\begin{split}
		\rmi\hbar\frac{\partial\phi_1(\bm{r})}{\partial t}=&\bigg(H_{\text{sp}}^1(\bm{r})+\tilde{U}_{11}\bigg\{\left[1-\frac{1}{N_{c_1}}\right]|\phi_1(\bm{r})|^2\\
		&+\frac{2}{N_{c_1}}\left<\tilde{\Lambda}_1^{\dagger}(\bm{r})\tilde{\Lambda}_1(\bm{r})\right>\bigg\}-\lambda_2^{1}\bigg)\phi_1(\bm{r})\\
		&-\frac{\tilde{U}_{11}}{N_{c_1}}\int|\phi_{1}(\bm{r}')|^2\Big[\left<\tilde{\Lambda}_1^{\dagger}(\bm{r}')\tilde{\Lambda}_1(\bm{r})\right>\phi_{1}(\bm{r}')\\
		&+\left<\tilde{\Lambda}_1(\bm{r}')\tilde{\Lambda}_1(\bm{r})\right>\phi_{1}^*(\bm{r}')\Big]\dr'\\
		&+\frac{\tilde{U}_{11}}{N_{c_1}}\left<\tilde{\Lambda}_1^2(\bm{r})\right>\phi_1^*(\bm{r}),
	\end{split}
\end{equation}
with nonlinear eigenvalue
\begin{equation}
	\begin{split}
	\lambda_2^1=&\int\Bigg[\phi_{1}^*(\bm{r})\Bigg(H_{\text{sp}}^{1}(\bm{r})+\tilde{U}_{11}\bigg\{\left[1-\frac{1}{N_{c_1}}\right]|\phi_{1}(\bm{r})|^2\\
	&+\frac{2}{N_{c_1}}\left<\tilde{\Lambda}_{1}^{\dagger}(\bm{r})\tilde{\Lambda}_{1}(\bm{r})\right>\bigg\}-\rmi\hbar\frac{\partial}{\partial t}\Bigg)\phi_{1}(\bm{r})
	\\&
	+\frac{\tilde{U}_{11}}{N_{c_1}}\left<\tilde{\Lambda}_{1}^2(\bm{r})\right>\phi_{1}^{*^2}(\bm{r})\Bigg]\dr.
	\end{split}
\end{equation}
\end{subequations}
This system of the GGP equation, MBdG equation and associated nonlinear eigenvalue is identical to (75), (63) and (73) respectively of \cite{gm}.

\subsection{Two-component condensates}

\subsubsection{Possible cases for two-component condensates}
Two-component condensates have been frequently realised in experiment (for example \cite{matthews,papp,durham}). They provide rich systems in which many different ground and excited states can exist, and have the potential to offer insights into instabilities \cite{itt,stkts} and the transition to turbulence in quantum systems \cite{tit,at}. In all possible cases, we have $n=2$ with $C=2$; i.e.\ either the two components are mutually coherent or they are incoherent. The three experimental realisations of two-component condensates that we have quoted above contain the three possible combinations of components, although from our formal point of view there are only two distinct cases. The first, Ref. \cite{matthews}, has realised a $^{87}$Rb condensate where the only difference between the two components is their internal spin state; the second, Ref. \cite{papp}, has realised a condensate with two isotopes ($^{85}$Rb and $^{87}$Rb) of the same atom, and the third, Ref. \cite{durham}, has realised a condensate with different atoms, $^{87}$Rb and $^{133}$Cs. Of these three experimental examples, the first one has coherent components, whereas the last two have incoherent components. When we consider each of these two cases in the following subsections, we will expand all summation terms for explicitness. 

\subsubsection{Mutually coherent components}

In the case when the two-components are mutually coherent, we have $l=1$ with $|p_1|=2$. In what follows in this section we define $p(1)=p(2)\equiv 1$ so that $\npk\equiv N_{c_1}$ and the lowest-order equations of motion are the coupled GP equations (\ref{order1}), which read
\begin{subequations}
	\label{order1_ex_2_1}
\begin{equation}
	\label{gp_ex_2_1}
	\begin{split}
		\rmi\hbar\frac{\partial\phi_k(\bm{r})}{\partial t}=&\Big[H_{\text{sp}}^k(\bm{r})+\Utkk|\phi_k(\bm{r})|^2
		+\tilde{U}_{12}|\phi_{3-k}(\bm{r})|^2
		\\&-\lambda_0^1\Big]\phi_k(\bm{r})+H_{\text{ob}}^{k(3-k)}(\bm{r})\phi_{3-k}(\bm{r}),
		\end{split}
\end{equation}
with nonlinear eigenvalue
\newpage
\begin{equation}
	\label{cp_1_ex_2_1}
	\begin{split}
	\lambda_0^1=&\int\Bigg\{\phi_{1}^*(\bm{r})\bigg[H_{\text{sp}}^{1}(\bm{r})+\tilde{U}_{11}|\phi_{1}(\bm{r})|^2+\tilde{U}_{12}|\phi_{2}(\bm{r})|^2
	\\&
	-\rmi\hbar\frac{\partial}{\partial t}\bigg]\phi_{1}(\bm{r})+\phi_{2}^*(\bm{r})\bigg[H_{\text{sp}}^{2}(\bm{r})+\tilde{U}_{22}|\phi_{2}(\bm{r})|^2
	\\&
	+\tilde{U}_{12}|\phi_{1}(\bm{r})|^2-\rmi\hbar\frac{\partial}{\partial t}\bigg]\phi_{2}(\bm{r})\\
		&+\phi_2^*(\bm{r})H_{\text{ob}}^{21}(\bm{r})\phi_{1}(\bm{r})+\phi_1^*(\bm{r})H_{\text{ob}}^{12}(\bm{r})\phi_{2}(\bm{r})\Bigg\}\dr,
	\end{split}
\end{equation}
\end{subequations}
where $k=1$, $2$.
\begin{widetext}

To next order we have the MBdG equations (\ref{mbdg}), that read (for $k=1$, $2$)
\begin{multline}
		\label{mbdg_ex_2_1}
		\rmi\hbar\frac{d\Lk}{dt}=\left[H_{\text{sp}}^k(\bm{r})+\tilde{U}_{kk}|\phi_k(\bm{r})|^2+\tilde{U}_{12}|\phi_{3-k}(\bm{r})|^2-\lambda_0^1\right]\Lk+H_{\text{ob}}^{k(3-k)}(\bm{r})\tilde{\Lambda}_{3-k}(\bm{r})+\int Q_{k1}(\bm{r},\bm{r}')\bigg\{\tilde{U}_{11}\left[\phi_1^*(\bm{r}')\tilde{\Lambda}_1(\bm{r}')+\text{H.c.}\right]\\
		+\tilde{U}_{12}\left[\phi_2^*(\bm{r}')\tilde{\Lambda}_2(\bm{r}')+\text{H.c.}\right]\bigg\}\phi_1(\bm{r}')+Q_{k2}(\bm{r},\bm{r}')\left\{\tilde{U}_{22}\left[\phi_2^*(\bm{r}')\tilde{\Lambda}_2(\bm{r}')+\text{H.c.}\right]+\tilde{U}_{12}\left[\phi_1^*(\bm{r}')\tilde{\Lambda}_1(\bm{r}')+\text{H.c.}\right]\right\}\phi_2(\bm{r}')\dr',
\end{multline}
coupled to the above GP equations (\ref{order1_ex_2_1}). Finally, our second order (consistent) dynamical equations of motion for the mutually coherent two-component condensate are given by these MBdG equations (\ref{mbdg_ex_2_1}) coupled to the GGP equation (\ref{ggp}):
	\begin{subequations}
	\label{ggp_ex_2_1}
\begin{multline}
	\label{ggp_ex_2_1a}
		\rmi\hbar\frac{\partial\phi_k(\bm{r})}{\partial t}=\Bigg(H_{\text{sp}}^k(\bm{r})+\tilde{U}_{kk}\left\{\left[1-\frac{1}{N_{c_{1}}}\right]|\phi_k(\bm{r})|^2+\frac{2}{N_{c_{1}}}\left<\Lkdag\Lk\right>\right\}+\tilde{U}_{12}\bigg\{\left[1-\frac{1}{N_{c_{1}}}\right]|\phi_{3-k}(\bm{r})|^2+\frac{1}{N_{c_{1}}}\left<\tilde{\Lambda}_{3-k}^{\dagger}(\bm{r})\tilde{\Lambda}_{3-k}(\bm{r})\right>\bigg\}\\
	\begin{split}
		&-\lambda_2^{1}\Bigg)\phi_k(\bm{r})+\frac{\tilde{U}_{kk}}{N_{c_{1}}}\left<\tilde{\Lambda}_{k}^2(\bm{r})\right>\phi_k^*(\bm{r})+\frac{\tilde{U}_{12}}{N_{c_{1}}}\left<\tilde{\Lambda}_{k}(\bm{r})\tilde{\Lambda}_{3-k}(\bm{r})\right>\phi_{3-k}^*(\bm{r})+\left[H_{\text{ob}}^{k(3-k)}(\bm{r})+\frac{\tilde{U}_{12}}{N_{c_{1}}}\left<\tilde{\Lambda}_{3-k}^{\dagger}(\bm{r})\tilde{\Lambda}_{k}(\bm{r})\right>\right]\phi_{3-k}(\bm{r})\\
		&-\frac{1}{N_{c_{1}}}\int\left[\tilde{U}_{11}|\phi_1(\bm{r}')|^2+\tilde{U}_{12}|\phi_2(\bm{r}')|^2\right]\left[\left<\tilde{\Lambda}_{1}^{\dagger}(\bm{r}')\tilde{\Lambda}_{k}(\bm{r})\right>\phi_1(\bm{r}')+\left<\tilde{\Lambda}_{1}(\bm{r}')\tilde{\Lambda}_{k}(\bm{r})\right>\phi_1^*(\bm{r}')\right]\\
	&+\left[\tilde{U}_{22}|\phi_2(\bm{r}')|^2+\tilde{U}_{12}|\phi_1(\bm{r}')|^2\right]\left[\left<\tilde{\Lambda}_{2}^{\dagger}(\bm{r}')\tilde{\Lambda}_{k}(\bm{r})\right>\phi_2(\bm{r}')+\left<\tilde{\Lambda}_{2}(\bm{r}')\tilde{\Lambda}_{k}(\bm{r})\right>\phi_2^*(\bm{r}')\right]\dr',	
	\end{split}
\end{multline}
with nonlinear eigenvalue
\begin{equation}
	\begin{split}
	\label{ggp_ex_2_1_lambda}
	\lambda_2^{1}=&\int\Bigg[\phi_{1}^*(\bm{r})\Bigg(H_{\text{sp}}^{1}(\bm{r})+\tilde{U}_{11}\left\{\left[1-\frac{1}{N_{c_1}}\right]|\phi_{1}(\bm{r})|^2+\frac{2}{N_{c_1}}\left<\tilde{\Lambda}_{1}^{\dagger}(\bm{r})\tilde{\Lambda}_{1}(\bm{r})\right>\right\}
	\\&	
	-\rmi\hbar\frac{\partial}{\partial t}
	+\tilde{U}_{12}\left\{\left[1-\frac{1}{N_{c_{1}}}\right]|\phi_2(\bm{r})|^2+\frac{1}{N_{c_1}}\left<\tilde{\Lambda}_{2}^{\dagger}(\bm{r})\tilde{\Lambda}_{2}(\bm{r})\right>\right\}\Bigg)\phi_{1}(\bm{r})\\
	&+\phi_{2}^*(\bm{r})\Bigg(H_{\text{sp}}^{2}(\bm{r})+\tilde{U}_{22}\left\{\left[1-\frac{1}{N_{c_1}}\right]|\phi_{2}(\bm{r})|^2+\frac{2}{N_{c_1}}\left<\tilde{\Lambda}_{2}^{\dagger}(\bm{r})\tilde{\Lambda}_{2}(\bm{r})\right>\right\}
	\\&
	-\rmi\hbar\frac{\partial}{\partial t}+\tilde{U}_{12}\left\{\left[1-\frac{1}{N_{c_{1}}}\right]|\phi_1(\bm{r})|^2+\frac{1}{N_{c_1}}\left<\tilde{\Lambda}_{1}^{\dagger}(\bm{r})\tilde{\Lambda}_{1}(\bm{r})\right>\right\}\Bigg)\phi_{2}(\bm{r})
	+\phi_{2}^*(\bm{r})H_{\text{ob}}^{21}(\bm{r})\phi_{1}(\bm{r})+\phi_{1}^*(\bm{r})H_{\text{ob}}^{12}(\bm{r})\phi_{2}(\bm{r})
	\\&+\frac{\tilde{U}_{12}}{{N_{c_1}}}
		\Bigg\{2\left<\tilde{\Lambda}_{1}(\bm{r})\tilde{\Lambda}_{2}(\bm{r})\right>\phi_1^*(\bm{r})\phi_{2}^*(\bm{r})+\left[\left<\tilde{\Lambda}_{2}^{\dagger}(\bm{r})\tilde{\Lambda}_{1}(\bm{r})\right>\phi_1^*(\bm{r})\phi_{2}(\bm{r})+\text{H.c.}\right]\Bigg\}\\
		&+\frac{1}{N_{c_1}}\bigg[\tilde{U}_{11}\left<\tilde{\Lambda}_{1}^2(\bm{r})\right>\phi_{1}^{*^2}(\bm{r})+\tilde{U}_{22}\left<\tilde{\Lambda}_{2}^2(\bm{r})\right>\phi_{2}^{*^2}(\bm{r})\bigg]\Bigg]\dr.
	\end{split}
\end{equation}
\end{subequations}

\end{widetext}


\subsubsection{Mutually incoherent components}

In the case 
\pagebreak
when the two-components are mutually incoherent, we have $l=2$ with $|p_1|=1$ and $|p_2|=1$. Within what follows in this section we define $p(1)\equiv 1$ and $p(2)\equiv 2$ so that $N_{c_{p(1)}}\equiv N_{c_1}$ and 
\pagebreak
$N_{c_{p(2)}}\equiv N_{c_2}$ (we will often write $N_{c_{p(3-k)}}\equiv N_{c_{3-k}}$ and $N_{c_{p(k)}}\equiv N_{c_k}$ as well).  The lowest order equations of motion are then  the coupled GP equations (\ref{order1}), which read
\begin{subequations}
	\label{order1_ex_2_2}
\begin{equation}
	\label{gp_ex_2_2}
	 \begin{split}
		\rmi\hbar\frac{\partial\phi_k(\bm{r})}{\partial t}=&\Bigg[H_{\text{sp}}^k(\bm{r})+\Utkk|\phi_k(\bm{r})|^2
		\\&+\tilde{U}_{12}\sqrt{\frac{\np3k}{\nk}}|\phi_{3-k}(\bm{r})|^2-\lambda_0^k\Bigg]\phi_k(\bm{r}),
	 \end{split}
\end{equation}
with nonlinear eigenvalues
\begin{equation}
	\label{cp_1_ex_2_2}
	\begin{split}
	\lambda_0^k&=\int\phi_k^*(\bm{r})\Bigg(H_{\text{sp}}^k(\bm{r})+\tilde{U}_{kk}|\phi_k(\bm{r})|^2\\
	&\quad+\tilde{U}_{12}\sqrt{\frac{\np3k}{\nk}}|\phi_{3-k}(\bm{r})|^2-\rmi\hbar\frac{\partial}{\partial t}\Bigg)\phi_k(\bm{r})\dr,
\end{split}
\end{equation}
\end{subequations}
where, as in the previous section, $k=1$, $2$. System (\ref{order1_ex_2_2}) describes a mutually incoherent two-component condensate in the thermodynamic limit at zero temperature. There have been extensive studies regarding this system in recent literature (see for example the review of \cite{tsubota} and a study on the ground and excited states in \cite{ma} as well as \cite{itt,tit,stkts,at}).

\begin{widetext}
To next order we have the MBdG equations (\ref{mbdg}), that read (for $k=1$, $2$)
\begin{equation}
	\label{mbdg_ex_2_2}
	\begin{split}
		\rmi\hbar\frac{d\Lk}{dt}=&\Bigg[H_{\text{sp}}^k(\bm{r})+\tilde{U}_{kk}|\phi_k(\bm{r})|^2+\tilde{U}_{12}\sqrt{\frac{N_{c_{3-k}}}{N_{c_k}}}|\phi_{3-k}(\bm{r})|^2
		-\lambda_0^{k}\Bigg]\Lk
		\\&
		+\int Q_{kk}(\bm{r},\bm{r}')\Bigg\{\tilde{U}_{kk}\bigg[\phi_k^*(\bm{r}')\tilde{\Lambda}_k(\bm{r}')
		+\text{H.c.}\bigg]\phi_k(\bm{r}')+\tilde{U}_{12}\bigg[\phi_{3-k}^*(\bm{r}')\tilde{\Lambda}_{3-k}(\bm{r}')
		+\text{H.c.}\bigg]\phi_k(\bm{r}')\Bigg\} \dr',
	\end{split}
\end{equation}
coupled to the above GP equations (\ref{order1_ex_2_2}). Finally, our second order (consistent) dynamical equations of motion for the mutually incoherent two-component condensate are given by these MBdG equations (\ref{order23_ex_2_2b}) coupled to the GGP equation (\ref{ggp}):
\begin{subequations}
	\label{order23_ex_2_2b}
\begin{equation}
	\label{ggp_ex_2_2}
	\begin{split}
		\rmi\hbar\frac{\partial\phi_k(\bm{r})}{\partial t}=&\Bigg(H_{\text{sp}}^k(\bm{r})+\tilde{U}_{kk}\Bigg\{\left[1-\frac{1}{N_{c_{k}}}\right]|\phi_k(\bm{r})|^2
		+\frac{2}{N_{c_{k}}}\left<\Lkdag\Lk\right>\Bigg\}\\
		&+\tilde{U}_{12}\sqrt{\frac{N_{c_{3-k}}}{N_{c_{k}}}}\bigg[|\phi_{3-k}(\bm{r})|^2+\frac{1}{N_{c_{3-k}}}\left<\tilde{\Lambda}_{3-k}^{\dagger}(\bm{r})\tilde{\Lambda}_{3-k}(\bm{r})\right>\bigg]
		-\lambda_2^{k}\Bigg)\phi_k(\bm{r})
		+\frac{\Utkk}{N_{c_{k}}}\left<\tilde{\Lambda}_k^2(\bm{r})\right>\phi_k^*(\bm{r})
		\\
		+&\frac{\tilde{U}_{12}}{N_{c_k}}\left[\left<\tilde{\Lambda}_{3-k}^{\dagger}(\bm{r})\tilde{\Lambda}_{k}(\bm{r})\right>\phi_{3-k}(\bm{r})+\left<\tilde{\Lambda}_{3-k}(\bm{r})\tilde{\Lambda}_{k}(\bm{r})\right>\phi^*_{3-k}(\bm{r})\right]\\
	&-\int\frac{|\phi_k(\bm{r}')|^2}{N_{c_k}}\Bigg\{\tilde{U}_{kk}\left[\left<\tilde{\Lambda}_{k}^{\dagger}(\bm{r}')\tilde{\Lambda}_{k}(\bm{r})\right>\phi_k(\bm{r}')+\left<\tilde{\Lambda}_{k}(\bm{r}')\tilde{\Lambda}_{k}(\bm{r})\right>\phi_k^*(\bm{r}')\right]\\
	&\quad+\tilde{U}_{12}\left[\left<\tilde{\Lambda}_{3-k}^{\dagger}(\bm{r}')\tilde{\Lambda}_{k}(\bm{r})\right>\phi_{3-k}(\bm{r}')+\left<\tilde{\Lambda}_{3-k}(\bm{r}')\tilde{\Lambda}_{k}(\bm{r})\right>\phi_{3-k}^*(\bm{r}')\right]\Bigg\}\dr',		
	\end{split}
\end{equation}
with nonlinear eigenvalues
\begin{equation}
	\begin{split}
	\lambda_2^{k}=&\int\phi_{k}^*(\bm{r})\Bigg(H_{\text{sp}}^{k}(\bm{r})+\tilde{U}_{kk}\Bigg\{\left[1-\frac{1}{N_{c_k}}\right]|\phi_{k}(\bm{r})|^2
	+\frac{2}{N_{c_k}}\left<\tilde{\Lambda}_{k}^{\dagger}(\bm{r})\tilde{\Lambda}_{k}(\bm{r})\right>\Bigg\}-\rmi\hbar\frac{\partial}{\partial t}
	\\&
	+\tilde{U}_{12}\sqrt{\frac{N_{c_{3-k}}}{N_{c_k}}}\bigg[|\phi_{3-k}(\bm{r})|^2
	+\frac{1}{N_{c_{3-k}}}\left<\tilde{\Lambda}_{3-k}^{\dagger}(\bm{r})\tilde{\Lambda}_{3-k}(\bm{r})\right>\bigg]\Bigg)\phi_{k}(\bm{r})+\frac{\tilde{U}_{kk}}{N_{c_k}}\left<\tilde{\Lambda}_{k}^2(\bm{r})\right>\phi_{k}^{*^2}(\bm{r})\\
+&\frac{\tilde{U}_{12}}{N_{c_k}}\left[\left<\tilde{\Lambda}_{3-k}^{\dagger}(\bm{r})\tilde{\Lambda}_{k}(\bm{r})\right>\phi_{3-k}(\bm{r})+\left<\tilde{\Lambda}_{3-k}(\bm{r})\tilde{\Lambda}_{k}(\bm{r})\right>\phi^*_{3-k}(\bm{r})\right]\phi_k^*(\bm{r})\dr,
	\end{split}
\end{equation}
\end{subequations}
\end{widetext}
again for $k=1$, $2$. 

\subsubsection{Comparison}

We provide a brief overview of the differences between the system of equations developed for the single component condensate [(\ref{mbdg_ex_1_1}) and (\ref{order23_ex_1_1b})] with those developed for the mutually coherent [(\ref{mbdg_ex_2_1}) and (\ref{ggp_ex_2_1})] and incoherent  [(\ref{mbdg_ex_2_2}) and (\ref{order23_ex_2_2b})] two-component condensates. 
Through a comparison of the GGP equation for the single component condensate, written in (\ref{ggp_ex_1}), with the GGP equations for the coherent or incoherent two-component condensate, written in (\ref{ggp_ex_2_1a}) and (\ref{ggp_ex_2_2}) respectively, the only additions that appear for the two-component condensate are those that involve $\tilde{U}_{12}$. These involve a density--density interaction term of type $\phi_{3-k}(\bm{r})$ and a fluctuation pair average of $\tilde{\Lambda}_{3-k}(\bm{r})$ and combinations thereof. This latter pair average term is always modified by a factor $(N_{c_{3-k}})^{-1}$ and so can be expected to be much smaller than the former condensate density--density term. 
Notice that the GGP equations for the two-component condensates, whether they are mutually coherent or mutually incoherent, differ only by the appearance of the one-body $H_{\text{ob}}$ term for the coherent case. This is in contrast to the MBdG equations of (\ref{mbdg_ex_2_1}) and (\ref{mbdg_ex_2_2}) which under the integrals contain the projectors $Q$, and it is here that one must recall their definition from Eq. (\ref{qdeq}), importantly the appearance of a Kroneker delta mapping term. This leads to two extra terms (although formally they are of the same form) in the case of the coherent components given in Eq. (\ref{mbdg_ex_2_1}) when compared to the case of the incoherent components given in Eq. (\ref{mbdg_ex_2_2}).




\subsection{Three-component condensates}

The last example is that of the three-component condensate. Whilst experimental realisations of such condensates are few,
the following example is worthy of inclusion since its represents a further component configuration not possible in the two-component condensate; in principle one will have a three-component system whenever there is a Rabi coupling within one of the species of a two-species condensate mixture experiment.  

A three-component condensate has $n=3$ with $C=3$; i.e.\ either the three components are mutually coherent or there are two mutually coherent components and one incoherent component or they are all mutually incoherent. We will only concentrate on the second of these three possibilities (the other two are straightforward generalisations of the two-component cases present above).  In the case when there are two mutually coherent components (say, component 1 and component 2) and one incoherent component (component 3), we have $l=2$ with $|p_1|=2$ and $|p_2|=1$. In what follows in this section we define $p(1)=p(2)\equiv 1$ and $p(3)\equiv 2$ so that $N_{c_{p(1)}}=N_{c_{p(2)}}\equiv N_{c_1}$ and $N_{c_{p(3)}}\equiv N_{c_2}$. The GP equations (\ref{order1}) then read
\begin{widetext}
\begin{subequations}
	\label{order1_ex_3_2}
\begin{equation}
	\label{gp_ex_3_2a}
		\rmi\hbar\frac{\partial\phi_k(\bm{r})}{\partial t}=\bigg[H_{\text{sp}}^k(\bm{r})+\tilde{U}_{k1}|\phi_{1}(\bm{r})|^2+\tilde{U}_{k2}|\phi_{2}(\bm{r})|^2+\sqrt{\frac{N_{c_2}}{N_{c_1}}}\tilde{U}_{k3}|\phi_{3}(\bm{r})|^2
		-\lambda_0^1\bigg]\phi_k(\bm{r})+H_{\text{ob}}^{k(3-k)}(\bm{r})\phi_{3-k}(\bm{r}),
\end{equation}
for $k=1$, $2$, and
\begin{equation}
	\label{gp_ex_3_2b}
		\rmi\hbar\frac{\partial\phi_3(\bm{r})}{\partial t}=\bigg\{H_{\text{sp}}^3(\bm{r})+\tilde{U}_{33}|\phi_{3}(\bm{r})|^2+\sqrt{\frac{N_{c_1}}{N_{c_2}}}\bigg[\tilde{U}_{13}|\phi_{1}(\bm{r})|^2+\tilde{U}_{23}|\phi_{2}(\bm{r})|^2\bigg]-\lambda_0^{2}\bigg\}\phi_3(\bm{r}),
\end{equation}
with nonlinear eigenvalues
\begin{equation}
	\label{cp_1_ex_3_2a}
	\begin{split}	
	\lambda_0^1=&\int\Bigg\{\phi_1^*(\bm{r})\bigg[H_{\text{sp}}^1(\bm{r})+\tilde{U}_{11}|\phi_{1}(\bm{r})|^2+\tilde{U}_{12}|\phi_{2}(\bm{r})|^2
	+\sqrt{\frac{N_{c_2}}{N_{c_1}}}\tilde{U}_{13}|\phi_{3}(\bm{r})|^2-\rmi\hbar\frac{\partial}{\partial t}\bigg]\phi_1(\bm{r})
	+\phi_2^*(\bm{r})\bigg[H_{\text{sp}}^2(\bm{r})
	\\&
	+\tilde{U}_{12}|\phi_{1}(\bm{r})|^2+\tilde{U}_{22}|\phi_{2}(\bm{r})|^2
	+\sqrt{\frac{N_{c_2}}{N_{c_1}}}\tilde{U}_{23}|\phi_{3}(\bm{r})|^2
	-\rmi\hbar\frac{\partial}{\partial t}\bigg]\phi_2(\bm{r})
	+\Big[\phi_2^*(\bm{r})H_{\text{ob}}^{21}(\bm{r})\phi_1(\bm{r})
	+\phi_1^*(\bm{r})H_{\text{ob}}^{12}(\bm{r})\phi_2(\bm{r})\Big]\Bigg\}\dr,
		\end{split}
\end{equation}
and
\begin{equation}
	\label{cp_1_ex_3_2b}
	\lambda_0^2=\int\phi_3^*(\bm{r})\bigg[H_{\text{sp}}^3(\bm{r})+\tilde{U}_{33}|\phi_{3}(\bm{r})|^2+\sqrt{\frac{N_{c_1}}{N_{c_2}}}\bigg(\tilde{U}_{13}|\phi_{1}(\bm{r})|^2
+\tilde{U}_{23}|\phi_{2}(\bm{r})|^2\bigg)-\rmi\hbar\frac{\partial}{\partial t}\bigg]\phi_3(\bm{r})\dr.
\end{equation}
\end{subequations}

To next order we have the MBdG equations (\ref{mbdg}), that read
\begin{subequations}
\label{order23_ex_3_2b}
\begin{equation}
	\label{mbdg_ex_3_2a}
	\begin{split}
		\rmi\hbar\frac{d\Lk}{dt}=&\bigg[H_{\text{sp}}^k(\bm{r})+\tilde{U}_{k1}|\phi_{1}(\bm{r})|^2+\tilde{U}_{k2}|\phi_{2}(\bm{r})|^2
		+\tilde{U}_{k3}\sqrt{\frac{N_{c_2}}{N_{c_1}}}|\phi_{3}(\bm{r})|^2-\lambda_0^1\bigg]\Lk+H_{\text{ob}}^{k(3-k)}(\bm{r})\tilde{\Lambda}_{3-k}(\bm{r})\\
		&+\int \Big[Q_{k1}(\bm{r},\bm{r}')\Big\{\tilde{U}_{11}\left[\phi_1^*(\bm{r}')\tilde{\Lambda}_1(\bm{r}')+\text{H.c.}\right]
		+\tilde{U}_{12}\left[\phi_2^*(\bm{r}')\tilde{\Lambda}_2(\bm{r}')+\text{H.c.}\right]
		+\tilde{U}_{13}\left[\phi_3^*(\bm{r}')\tilde{\Lambda}_3(\bm{r}')+\text{H.c.}\right]\Big\}\phi_1(\bm{r}')\\
		&+Q_{k2}(\bm{r},\bm{r}')\Big\{\tilde{U}_{22}\left[\phi_2^*(\bm{r}')\tilde{\Lambda}_2(\bm{r}')+\text{H.c.}\right]
		+\tilde{U}_{12}\left[\phi_1^*(\bm{r}')\tilde{\Lambda}_1(\bm{r}')+\text{H.c.}\right]
		+\tilde{U}_{23}\left[\phi_3^*(\bm{r}')\tilde{\Lambda}_3(\bm{r}')+\text{H.c.}\right]\Big\}\phi_2(\bm{r}')\Big]\dr',
	\end{split}
\end{equation}
for $k=1$, $2$, and
\begin{multline}
	\label{mbdg_ex_3_2b}
		\rmi\hbar\frac{d\tilde{\Lambda}_3(\bm{r})}{dt}=\Bigg\{H_{\text{sp}}^3(\bm{r})+\tilde{U}_{33}|\phi_{3}(\bm{r})|^2+\sqrt{\frac{N_{c_1}}{N_{c_2}}}\bigg[\tilde{U}_{13}|\phi_{1}(\bm{r})|^2
		+\tilde{U}_{23}|\phi_{2}(\bm{r})|^2\bigg]-\lambda_0^{2}\Bigg\}\tilde{\Lambda}_3(\bm{r})
		\\
		+\int Q_{33}(\bm{r},\bm{r}')\Big\{\tilde{U}_{33}\left[\phi_3^*(\bm{r}')\tilde{\Lambda}_3(\bm{r}')+\text{H.c.}\right]
		+\tilde{U}_{13}\left[\phi_1^*(\bm{r}')\tilde{\Lambda}_1(\bm{r}')+\text{H.c.}\right]
		+\tilde{U}_{23}\left[\phi_2^*(\bm{r}')\tilde{\Lambda}_2(\bm{r}')+\text{H.c.}\right]\Big\}\phi_3(\bm{r}')\dr',
\end{multline}
\end{subequations}
coupled to the above GP equations (\ref{order1_ex_3_2}). Finally, our second order (consistent) dynamical equations of motion for the three-component condensate are given by these MBdG equations (\ref{order23_ex_3_2b}) coupled to the GGP equation (\ref{ggp}):
\begin{subequations}
\begin{multline}
	\label{ggp_ex_3_2a}
		\rmi\hbar\frac{\partial\phi_k(\bm{r})}{\partial t}=\Bigg(H_{\text{sp}}^k(\bm{r})+\tilde{U}_{kk}\bigg\{\left[1-\frac{1}{N_{c_1}}\right]|\phi_k(\bm{r})|^2
		+2\frac{\left<\Lkdag\Lk\right>}{{N_{c_1}}}\bigg\}-\lambda_2^{1}+\tilde{U}_{12}\bigg\{\left[1-\frac{1}{N_{c_1}}\right]|\phi_{3-k}(\bm{r})|^2
		+\frac{\left<\tilde{\Lambda}_{3-k}^{\dagger}(\bm{r})\tilde{\Lambda}_{3-k}(\bm{r})\right>}{N_{c_1}}\bigg\}
	\\
	\begin{split}	
	&
	+\sqrt{\frac{{N_{c_2}}}{{N_{c_1}}}}\tilde{U}_{k3}\left(|\phi_3(\bm{r})|^2+\frac{\left<\tilde{\Lambda}_3^{\dagger}(\bm{r})\tilde{\Lambda}_3(\bm{r})\right>}{{N_{c_2}}}\right)\Bigg)\phi_k(\bm{r})
	+\frac{\Utkk}{{N_{c_1}}}\left<\tilde{\Lambda}_k^2(\bm{r})\right>\phi_k^*(\bm{r})
	+H_{\text{ob}}^{k(3-k)}(\bm{r})\phi_{3-k}(\bm{r})\\
	&+\frac{\tilde{U}_{12}}{N_{c_1}}\Big[\left<\tilde{\Lambda}_{3-k}^{\dagger}(\bm{r})\tilde{\Lambda}_{k}(\bm{r})\right>\phi_{3-k}(\bm{r})
	+\left<\tilde{\Lambda}_{3-k}(\bm{r})\tilde{\Lambda}_{k}(\bm{r})\right>\phi_{3-k}^*(\bm{r})\Big]+\frac{\tilde{U}_{k3}}{N_{c_1}}\Big[\left<\tilde{\Lambda}_{3}^{\dagger}(\bm{r})\tilde{\Lambda}_{k}(\bm{r})\right>\phi_{3}(\bm{r})
	+\left<\tilde{\Lambda}_{3}(\bm{r})\tilde{\Lambda}_{k}(\bm{r})\right>\phi_{3}^*(\bm{r})\Big]\\	
&-\int \Bigg\{\left[\frac{\tilde{U}_{11}}{N_{c_1}}|\phi_1(\bm{r}')|^2+\frac{\tilde{U}_{12}}{N_{c_1}}|\phi_2(\bm{r}')|^2
\right]
\left[\left<\tilde{\Lambda}_{1}^{\dagger}(\bm{r}')\tilde{\Lambda}_{k}(\bm{r})\right>\phi_1(\bm{r}')+\left<\tilde{\Lambda}_{1}(\bm{r}')\tilde{\Lambda}_{k}(\bm{r})\right>\phi_1^*(\bm{r}')\right]\\
	&\quad+\left[\frac{\tilde{U}_{12}}{N_{c_1}}|\phi_1(\bm{r}')|^2+\frac{\tilde{U}_{22}}{N_{c_1}}|\phi_2(\bm{r}')|^2
	\right]
	\left[\left<\tilde{\Lambda}_{2}^{\dagger}(\bm{r}')\tilde{\Lambda}_{k}(\bm{r})\right>\phi_2(\bm{r}')+\left<\tilde{\Lambda}_{2}(\bm{r}')\tilde{\Lambda}_{k}(\bm{r})\right>\phi_2^*(\bm{r}')\right]\\
	&\quad+\left[\frac{\tilde{U}_{13}}{N_{c_1}}|\phi_1(\bm{r}')|^2+\frac{\tilde{U}_{23}}{N_{c_1}}|\phi_2(\bm{r}')|^2
	\right]
	\left[\left<\tilde{\Lambda}_{3}^{\dagger}(\bm{r}')\tilde{\Lambda}_{k}(\bm{r})\right>\phi_3(\bm{r}')+\left<\tilde{\Lambda}_{3}(\bm{r}')\tilde{\Lambda}_{k}(\bm{r})\right>\phi_3^*(\bm{r}')\right]\Bigg\}\dr',		
	\end{split}
\end{multline}
for $k=1$, $2$, and
\begin{equation}
	\label{ggp_ex_3_2b}
	\begin{split}
		\rmi\hbar\frac{\partial\phi_3(\bm{r})}{\partial t}=&\Bigg(H_{\text{sp}}^3(\bm{r})+\tilde{U}_{33}\Bigg\{\left[1-\frac{1}{N_{c_2}}\right]|\phi_3(\bm{r})|^2
+\frac{2}{N_{c_2}}\left<\tilde{\Lambda}_3^{\dagger}(\bm{r})\tilde{\Lambda}_3(\bm{r})\right>\Bigg\}-\lambda_2^{2}+\sqrt{\frac{N_{c_1}}{N_{c_2}}}\bigg[\tilde{U}_{13}|\phi_{1}(\bm{r})|^2\\
		&+\tilde{U}_{23}|\phi_{2}(\bm{r})|^2+\frac{1}{N_{c_{1}}}\left<\tilde{\Lambda}_{1}^{\dagger}(\bm{r})\tilde{\Lambda}_{1}(\bm{r})\right>
		+\frac{1}{N_{c_{1}}}\left<\tilde{\Lambda}_{2}^{\dagger}(\bm{r})\tilde{\Lambda}_{2}(\bm{r})\right>\bigg]\Bigg)\phi_3(\bm{r})+\frac{\tilde{U}_{33}}{N_{c_2}}\left<\tilde{\Lambda}_3^2(\bm{r})\right>\phi_3^*(\bm{r})\\
	&+\frac{\tilde{U}_{13}}{N_{c_2}}\Big[\left<\tilde{\Lambda}_{1}^{\dagger}(\bm{r})\tilde{\Lambda}_{3}(\bm{r})\right>\phi_{1}(\bm{r})
	+\left<\tilde{\Lambda}_{1}(\bm{r})\tilde{\Lambda}_{3}(\bm{r})\right>\phi_{1}^*(\bm{r})\Big]+\frac{\tilde{U}_{23}}{N_{c_2}}\Big[\left<\tilde{\Lambda}_{2}^{\dagger}(\bm{r})\tilde{\Lambda}_{3}(\bm{r})\right>\phi_{2}(\bm{r})
	+\left<\tilde{\Lambda}_{2}(\bm{r})\tilde{\Lambda}_{3}(\bm{r})\right>\phi_{2}^*(\bm{r})\Big]\\		
	&-\int\frac{|\phi_3(\bm{r}')|^2}{N_{c_2}}\Bigg\{\tilde{U}_{33}\left[\left<\tilde{\Lambda}_3^{\dagger}(\bm{r}')\tilde{\Lambda}_3(\bm{r})\right>\phi_3(\bm{r}')+\left<\tilde{\Lambda}_3(\bm{r}')\tilde{\Lambda}_3(\bm{r})\right>\phi_3^*(\bm{r'})\right]\\
	&+\tilde{U}_{13}\Big[\left<\tilde{\Lambda}_{1}^{\dagger}(\bm{r}')\tilde{\Lambda}_{3}(\bm{r})\right>\phi_{1}(\bm{r}')
	+\left<\tilde{\Lambda}_{1}(\bm{r}')\tilde{\Lambda}_{3}(\bm{r})\right>\phi_{1}^*(\bm{r}')\Big]+\tilde{U}_{23}\Big[\left<\tilde{\Lambda}_{2}^{\dagger}(\bm{r}')\tilde{\Lambda}_{3}(\bm{r})\right>\phi_{2}(\bm{r}')
	+\left<\tilde{\Lambda}_{2}(\bm{r}')\tilde{\Lambda}_{3}(\bm{r})\right>\phi_{2}^*(\bm{r}')\Big]\Bigg\}\dr',	
	\end{split}
\end{equation}
\end{subequations}
\end{widetext}
where the nonlinear eigenvalues $\lambda_{2}^{1}$ and $\lambda_{2}^{2}$ follow from (\ref{lam222}).

\section{Conclusions}
\label{concs}

We have extended the number-conserving formalism developed for single-component Bose--Einstein condensed systems in \cite{gm} to general multi-component configurations. In the number-conserving approach, the system of equations are derived from approximations to the governing Hamiltonian, written generally to encompass any number of components, in a wide variety of mutually coherent and incoherent configurations.  Differing levels of approximation in the Hamiltonian generate dynamics described by multi-component equivalents to the Gross--Piteavskii equation, the number-conserving modified Bogoliubov--de Gennes equations (\ref{mbdg}) coupled to the Gross--Pitaevskii equation, and the generalised Gross--Pitaevskii equation (\ref{ggp}) coupled to the modified Bolgoliubov--de Gennes equations.  The last of these yields self-consistent equations of motion with regard to the condensate and non-condensate number.  We have looked specifically at two-component condensates, both where the components are mutually coherent (two internal states of the same atom), and where they are mutually incoherent (most obviously two completely different atomic species), and at a three-component configuration where two of the components are mutually coherent with respect to each other, but not with respect to a third component (e.g., two internal states of one species of atom, and another species of atom, all within the same experimental setup).  This provides the essential framework for detailed study of specific multi-component condensate configurations, within a canonical, or number-conserving, formalism.


\section*{Acknowledgments}
We thank A. S. Bradley, T. P. Billam, N. P. Proukakis, and M. J. Edmonds for stimulating discussions.  The authors wish to acknowledge funding support from Marie Curie Fellowship NUM2BEC (Grant No. 300285) and from EPSRC (Grant No. EP/K03250X/1),  and The Royal Society (Grant No. IE110202). P.M. also acknowledges financial support from the Fondation des Treilles.

\appendix
\label{appx}

\section{Reformulated Hamiltonian}
\label{reform}

Our $n$-component condensate is described by $n$ field operators, one for each component. The effective Hamiltonian is then given by (Eq.  (\ref{Ham}), rewritten here for convenience),
\begin{equation}
	\label{Ham_app}
	\begin{split}
	\hat{H}(t)=&\int\sk1n\Big[\Pkd\Hk\Pk\\
	&+\frac{\Ukk}{2}\Pkd\Pkd\Pk\Pk\Big]\dr\\
	&+\int\sum_{\substack{j,k\\j< k}}^n\Ujk\Pjd\Pkd\Pj\Pk\dr\\
	&+\int\sum_{\substack{j,k\\j\neq k}}^n\Pjd H_{\text{ob}}^{jk}(\bm{r},t)\Pk\dr.
	\end{split}
\end{equation}
In order to obtain a set of dynamical equations to describe the $n$-component condensate, we must, as a first step, reformulate this effective Hamiltonian in terms of the fluctuation operators (\ref{fluc}). In this appendix we show the details to obtain this exact reformulation. 

We first write each of the field operators in terms of the condensate and noncondensate parts (\ref{cond}): $\Pk=\hat{a}_{c_{p(k)}}(t)\phi_k(\bm{r},t)+\delta\hat{\Psi}_k(\bm{r},t)$. This gives us an effective Hamiltonian that is given in terms of the annihilation (and creation) operators of the condensate and noncondensate parts. By suitable rearrangement of each of these terms, using the commutator relations of (\ref{commutator}) and  ${\hat{N}_{c_{p(k)}}(t)}\equiv\hat{a}^{\dagger}_{c_{p(k)}}(t)\hat{a}_{c_{p(k)}}(t)$, we are able to write each of these terms solely as a product of annihilation operators and noncondensate parts (or equivalent Hermitian conjugates). This means we can replace each of these products with a fluctuation operator (\ref{fluc}). 

After collecting terms in products of $\tilde{\Lambda}$, this exact reformulation gives $\hat{H}=\hat{H}^{(\tilde{\Lambda}^0)}+\hat{H}^{(\tilde{\Lambda}^1)}+\hat{H}^{(\tilde{\Lambda}^2)}+\hat{H}^{(\tilde{\Lambda}^3)}+\hat{H}^{(\tilde{\Lambda}^4)}$, where
\begin{subequations}
	\label{Hamil}
\begin{equation}
	\begin{split}
	\hat{H}^{(\tilde{\Lambda}^0)}=&\int\sk1n\Bigg\{\nhpk\bigg[\phi_k^*(\bm{r})H_{\text{sp}}^k(\bm{r})\phi_k(\bm{r})
	\\&
	+\frac{(\nhpk-1)}{\npk}\frac{\Utkk}{2}|\phi_k(\bm{r})|^4\bigg]\\
	&+\sjk\Utjk\frac{\nhpj(\nhpk-\delta_{j,k}^p)}{\sqrt{\npj\npk}}|\phi_j(\bm{r})|^2|\phi_k(\bm{r})|^2\\
	&+\sum_{\substack{j=1\\j\neq k}}^n\nhpk\phi_j^*(\bm{r})H_{\text{ob}}^{jk}(\bm{r})\phi_k(\bm{r})\Bigg\}\dr,
\end{split}
\end{equation}
\begin{equation}
	\begin{split}
	\hat{H}^{(\tilde{\Lambda}^1)}=&\int\sk1n\Bigg\{\sqrt{\npk}\left[\phi_k^*(\bm{r})H_{\text{sp}}^k(\bm{r})\Lk+\text{H.c.}\right]\\
	&+\Utkk\left\{\phi_k^*(\bm{r})\frac{[\nhpk-1]}{\sqrt{\npk}}\Lk+\text{H.c.}\right\}|\phi_k(\bm{r})|^2\\
	&+\sum_{\substack{j=1\\j\neq k}}^n\Utjk\bigg\{\phi_j^*(\bm{r})\frac{[\nhpk-\delta_{j,k}^p]}{\sqrt{\npk}}\Lj+\text{H.c.}\bigg\}|\phi_k(\bm{r})|^2\\
	&+\sum_{\substack{j=1\\j\neq k}}^n\sqrt{\npk}\bigg[\phi_j^*(\bm{r})H_{\text{ob}}^{jk}(\bm{r})\Lk
	\\&
	+\Ljdag H_{\text{ob}}^{jk}(\bm{r})\phi_k(\bm{r})\bigg]\Bigg\}\dr,
	\end{split}
\end{equation}
\begin{equation}
	\begin{split}
	\hat{H}^{(\tilde{\Lambda}^2)}=&\int\sk1n\Bigg\{\Lkdag\bigg[\frac{\npk}{\nhpk}H_{\text{sp}}^k(\bm{r})+\frac{2\Utkk(\nhpk-1)}{\nhpk}\\
	&\times |\phi_k(\bm{r})|^2\bigg]\Lk+\frac{\Utkk}{2}\left[\phi_k^{*^2}(\bm{r})\tilde{\Lambda}_k^2(\bm{r})+\text{H.c.}\right]\\
	&+\sum_{\substack{j=1\\j\neq k}}^n\Ljdag\frac{\npk}{\nhpk}H_{\text{ob}}^{jk}(\bm{r})\Lk\\
	&+\sum_{\substack{j=1\\j\neq k}}^n\Utjk\Bigg\{\frac{1}{2}\left[\phi_j^*(\bm{r})\phi_k^*(\bm{r})\Lj\Lk+\text{H.c.}\right]\\
	&+\Ljdag\left(\nhpk-\delta_{j,k}^p\right)\bigg[\frac{1}{{\nhpj}}\sqrt{\frac{\npj}{\npk}}\\
	&\times\phi_k(\bm{r})\Lj+\frac{1}{\nhpk}\phi_j(\bm{r})\Lk\bigg]\phi_k^*(\bm{r})\Bigg\}\Bigg\}\dr,
	\end{split}
\end{equation}
\begin{equation}
	\begin{split}
	\hat{H}^{(\tilde{\Lambda}^3)}=&\int\sk1n\Bigg\{\Utkk\left[\phi_k^*(\bm{r})\Lkdag\frac{\sqrt{\npk}}{\nhpk}\tilde{\Lambda}_k^2(\bm{r})+\text{H.c.}\right]\\
	&+\sum_{\substack{j=1\\j\neq k}}^n\Utjk\bigg[\phi_j^*(\bm{r})\Lkdag\frac{\sqrt{\npk}}{\nhpk}\Lj\Lk+\text{H.c.}\bigg]\Bigg\}\dr,
	\end{split}
\end{equation}
\begin{equation}
	\begin{split}
	\hat{H}^{(\tilde{\Lambda}^4)}=&\int\sk1n\bigg[\frac{\Utkk}{2}\tilde{\Lambda}^{\dagger^2}_k(\bm{r})\frac{\npk}{\nhpk\left(\nhpk-1\right)}\tilde{\Lambda}_k^2(\bm{r})\\
	&+\sjk\Utjk\frac{\delta\hat{\Psi}_j^{\dagger}(\bm{r})\delta\hat{\Psi}_k^{\dagger}(\bm{r})\delta\hat{\Psi}_j(\bm{r})\delta\hat{\Psi}_k(\bm{r})}{\sqrt{\npj\npk}}\bigg]\dr.
\end{split}
\end{equation}
\end{subequations}

\section{Approximation to number and fluctuation operators}
\label{app_1}

The number fluctuations of the condensate and noncondensate components within each subset must be equal and opposite, i.e.
 \begin{equation}
	\begin{split}
	\nhpk=&\npk
	+\sum_{k'=1}^{n}\delta_{k,k'}^p\int \biggl[ \left<\delta\hat{\Psi}^{\dagger}_{k'}(\bm{r})\delta\hat{\Psi}_{k'}(\bm{r})\right>
	\\&
	-\delta\hat{\Psi}^{\dagger}_{k'}(\bm{r})\delta\hat{\Psi}_{k'}(\bm{r})\biggr]\dr\\
	=&\npk+\sum_{k'=1}^{n}\delta_{k,k'}^p\int \Biggl[ \left<\tilde{\Lambda}^{\dagger}_{k'}(\bm{r})\frac{\npkd}{\nhpkd}\tilde{\Lambda}_{k'}(\bm{r})\right>\\
	&-\tilde{\Lambda}^{\dagger}_{k'}(\bm{r})\frac{\npkd}{\nhpkd}\tilde{\Lambda}_{k'}(\bm{r})\Biggr]\dr,
	\label{n_app}
\end{split}
\end{equation}
where we have used (\ref{lams}). To zeroth- (and first-) order in the fluctuation operators, $\nhpk=\npk$, whereas to second-order we have
\begin{equation}
	\label{numapp_app}
	\nhpk=\npk+\sum_{k'=1}^{n}\delta_{k,k'}^p\int \biggl[ \left<\tilde{\Lambda}^{\dagger}_{k'}(\bm{r})\tilde{\Lambda}_{k'}(\bm{r})\right>
	-\tilde{\Lambda}^{\dagger}_{k'}(\bm{r})\tilde{\Lambda}_{k'}(\bm{r})\biggr] \dr.
\end{equation}

We can now use (\ref{numapp_app}) to express the commutation relation $[\tilde{\Lambda}_k(\bm{r}), \tilde{\Lambda}_{k'}^{\dagger}(\bm{r}')]$ (\ref{comm}) in terms of the condensate numbers and expectation values of $\tilde{\Lambda}_k(\bm{r})$ and $\tilde{\Lambda}_{k'}^{\dagger}(\bm{r}')$: from (\ref{lams}), (\ref{comm})  and (\ref{numapp_app}) we have
\begin{multline}
[\Lk,\Lkddd] \approx Q_{kk'}(\bm{r},\bm{r}')\Bigg\{1 + \sum_{k''=1}^{n}\delta_{k,k''}^p
\\ 
\times
\int \Bigg[\frac{\left<\tilde{\Lambda}_{k''}^{\dagger}(\bm{r}'')\tilde{\Lambda}_{k''}(\bm{r}'')\right>}{\npk}
-\frac{\tilde{\Lambda}_{k''}^{\dagger}(\bm{r}'')\tilde{\Lambda}_{k''}(\bm{r}'')}{\npk}\Bigg]\drdd\Bigg\}
\\
-\frac{\delta_{k,k'}^p}{(1+\nhpk)}\Lkddd\Lk,
\end{multline}
where we have written $\nhpk=\nhpkd$ and $\npk=\npkd$. We may replace the $(1+\nhpk)^{-1}$ term by $\npk^{-1}$, as the resulting difference will only be to quartic order. Finally, to a Gaussian level of approximation we may replace pairwise products of the fluctuation operators $\tilde{\Lambda}(\bm{r})$ and $\tilde{\Lambda}^{\dagger}(\bm{r})$ by their expectation values \cite{gm}. We thus write
\begin{equation}
	\begin{split}
[\Lk,\Lkddd]&\approx Q_{kk'}(\bm{r},\bm{r}')-\frac{\left<\Lkddd\Lk\right>}{\npk}\delta_{k,k'}^p,
\end{split}
\label{flucapp_app}
\end{equation}
whereas to zeroth and first order, the commutator may be approximated by 
\begin{equation}
[\Lk,\Lkddd]\approx Q_{kk'}(\bm{r},\bm{r}').
\label{fluccomzero_app}
\end{equation}

\section{Cubic Hamiltonian}
\label{reform_cubic}

The effective Hamiltonian, written in terms of fluctuation operators in (\ref{Hamil}), is an exact reformulation of the effective Hamiltonian written in terms of the field operators (\ref{Ham}). To make progress with the reformulated Hamiltonian of (\ref{Hamil}) we must consistently deal with the terms of cubic and quartic order in $\tilde{\Lambda}$. To achieve this, we need to use the approximation to the number operators $\hat{N}_{c_{p(k)}}$, given in (\ref{numapp}). Note that, while (\ref{numapp}) is a second-order approximation to the number operators, this is sufficient to retain consistency. We can then substitute (\ref{numapp}) into (\ref{Hamil}), expand any terms cubic in the fluctuation operators according to the Hartree--Fock factorisation (\ref{htf}), and neglect terms quartic in the fluctuation operators. Our third-order approximation to the full Hamiltonian (\ref{Hamil}) is then found to be given by (after collecting terms in products of $N_c$) $\hat{H}_3=\hat{H}_3^{(N_c^1)}+\hat{H}_3^{(N_c^{1/2})}+\hat{H}_3^{(N_c^0)}+\hat{H}_3^{(N_c^{-1/2})}$, where we further split $\hat{H}_3^{(N_c^0)}=\hat{H}_3^{(N_c^0)_{a}}+\hat{H}_3^{(N_c^0)_{b}}$ and $\hat{H}_3^{(N_c^{-1/2})}=\hat{H}_3^{(N_c^{-1/2})_a}+\hat{H}_3^{(N_c^{-1/2})_b}$. These expressions are 
\begin{subequations}
	\label{Ham3_1}
\begin{equation}
	\begin{split}
	\hat{H}_3^{(N_c^1)}=&\int\sk1n\Bigg\{\npk\phi_k^*(\bm{r})\left[H_{\text{sp}}^k(\bm{r})+\frac{\Utkk}{2}|\phi_k(\bm{r})|^2\right]\phi_k(\bm{r})\\
	&+\sjk\Utjk\sqrt{\npj\npk}|\phi_j(\bm{r})|^2|\phi_k(\bm{r})|^2\\
	&+\sum_{\substack{j=1\\j\neq k}}^n\npk\phi_j^*(\bm{r})H_{\text{ob}}^{jk}(\bm{r})\phi_k(\bm{r})\Bigg\}\dr,
\end{split}
\end{equation}
\begin{equation}
	\begin{split}
	\hat{H}_3^{(N_c^{1/2})}=&\int\sk1n\Bigg(\sqrt{\npk}\bigg\{\phi_k^*(\bm{r})\bigg[H_{\text{sp}}^k(\bm{r})+\Utkk|\phi_k(\bm{r})|^2\bigg]\Lk\\
	&+\text{H.c.}\bigg\}+\sum_{\substack{j=1\\j\neq k}}^n\Utjk\sqrt{\npk}\left[\phi_j^*(\bm{r})\Lj+\text{H.c.}\right]|\phi_k(\bm{r})|^2\\
	&+\sum_{\substack{j=1\\j\neq k}}^n\sqrt{\npk}\bigg[\phi_j^*(\bm{r})H_{\text{ob}}^{jk}(\bm{r})\Lk
	\\&
	+\Ljdag H_{\text{ob}}^{jk}(\bm{r})\phi_k(\bm{r})\bigg]\Bigg)\dr,
	\end{split}
\end{equation}
\begin{equation}
	\begin{split}
	\hat{H}_3^{(N_c^0)_{a}}=&\int\sk1n\Bigg(\Lkdag\left[H_{\text{sp}}^k(\bm{r})+2\Utkk|\phi_k(\bm{r})|^2\right]\Lk\\
	&
	+\frac{\Utkk}{2}\left[\phi_k^{*^2}(\bm{r})\tilde{\Lambda}_k^2(\bm{r})+\text{H.c.}\right]\\
	&+\sum_{\substack{j=1\\j\neq k}}^n\Utjk\Bigg\{\frac{1}{2}\left[\Lj\Lk\phi_j^*(\bm{r})\phi_k^*(\bm{r})+\text{H.c.}\right]\\
	&+\Ljdag\Bigg[\Lk\phi_j(\bm{r})+\Lj\phi_k(\bm{r})\\
	&\times\sqrt{\frac{\npk}{\npj}}\Bigg]\phi_k^*(\bm{r})\Bigg\}+\sum_{\substack{j=1\\j\neq k}}^n\Ljdag H_{\text{ob}}^{jk}(\bm{r})\Lk\Bigg)\dr\\
	&-\int\sk1n\Bigg[\frac{\Utkk}{2}|\phi_k(\bm{r})|^4
	\\&
	+\sjk\delta_{j,k}^p\Utjk|\phi_j(\bm{r})|^2|\phi_k(\bm{r})|^2\Bigg]\dr,
	\end{split}
\end{equation}
\begin{equation}
	\begin{split}	
	\hat{H}_3^{(N_c^0)_{b}}=&\int\sk1n\Bigg(\phi_k^*(\bm{r})\left[H_{\text{sp}}^k(\bm{r})+\Utkk|\phi_k(\bm{r})|^2\right]\phi_k(\bm{r})\dr\\
	&\times\sum_{k'=1}^{n}\delta_{k,k'}^p\int\bigg[\left<\Lkddd\Lkdash\right>-\Lkddd\Lkdash\dr'\bigg]\\
	&+\int\sum_{\substack{j=1\\j\neq k}}^n\Utjk|\phi_j(\bm{r})|^2|\phi_k(\bm{r})|^2\dr\\
	&\times\sqrt{\frac{\npj}{\npk}}\sum_{k'=1}^{n}\delta_{k,k'}^p\int\bigg[\left<\Lkddd\Lkdash\right>\\
	&-\Lkddd\Lkdash\bigg]\dr'\\
	&+\int\sum_{\substack{j=1\\j\neq k}}^n\phi_j^*(\bm{r})H_{\text{ob}}^{jk}(\bm{r})\phi_k(\bm{r})\dr\\
	&\times\sum_{k'=1}^{n}\delta_{k,k'}^p\int
	\left[
	\left<\Lkddd\Lkdash\right>-\Lkddd\Lkdash\right]\Bigg)\dr',
\end{split}
\end{equation}
\newpage
\begin{equation}
	\begin{split}
	\hat{H}_3^{(N_c^{-1/2})_{a}}=&\int\sk1n\Bigg(\frac{\Utkk}{\sqrt{\npk}}\Bigg\{\phi_k^*(\bm{r})\bigg[2\left<\Lkdag\Lk\right>\\
	&\times\Lk+\Lkdag\left<\tilde{\Lambda}_k^2(\bm{r})\right>\bigg]+\text{H.c.}\Bigg\}\\
	&+\sum_{\substack{j=1\\j\neq k}}^n\Utjk\Bigg\{\frac{\phi_k^*(\bm{r})}{\sqrt{\npj}}\Bigg[\left<\Ljdag\Lj\right>\Lk\\
	&+\left<\Ljdag\Lk\right>\Lj+\left<\Lj\Lk\right>\Ljdag\Bigg]\\
	&+\text{H.c.}\Bigg\}\Bigg)\dr-\int\sk1n\Bigg\{\frac{\Utkk}{\sqrt{\npk}}\\
	&\times\left[\phi_k^*(\bm{r})\Lk+\text{H.c.}\right]|\phi_k(\bm{r})|^2\\
	&+\sum_{\substack{j=1\\j\neq k}}^n\delta_{j,k}^p\frac{\Utjk}{\sqrt{\npk}}\left[\phi_j^*(\bm{r})\Lj+\text{H.c.}\right]|\phi_k(\bm{r})|^2\Bigg\}\dr,
\end{split}
\end{equation}
\newpage
\begin{equation}
	\begin{split}	
	\hat{H}_3^{(N_c^{-1/2})_{b}}=&-\int\sk1n\Bigg(\frac{\Utkk}{\sqrt{\npk}}|\phi_k(\bm{r})|^2\sum_{k'=1}^{n}\delta_{k,k'}^p\\
	&\times\int\Bigg\{\phi_k^*(\bm{r})\Big[\left<\Lkddd\Lk\right>\Lkdash\\
	&+\left<\Lkdash\Lk\right>\Lkddd\Big]+\text{H.c.}\Bigg\}\dr'\\
	&+\sum_{\substack{j=1\\j\neq k}}^n\Utjk\frac{|\phi_k(\bm{r})|^2}{\sqrt{\npk}}\sum_{k'=1}^{n}\delta_{k,k'}^p\\
	&\times \int\bigg\{\phi_j^*(\bm{r})\Big[\left<\Lkddd\Lj\right>\Lkdash\\
	&+\left<\Lkdash\Lj\right>\Lkddd\Big]+\text{H.c.}\bigg\}\dr'\Bigg)\dr.
\end{split}
\end{equation}
\end{subequations}

\section{Evolution equations}
\label{app_2}

In general the Heisenberg time evolution of the fluctuation operators is given by
\begin{equation}
	\label{lambda_time_app}
	\rmi\hbar\frac{d}{dt}\Lk=[\Lk,\hat{H}]+\rmi\hbar\frac{\partial}{\partial t}\Lk,
\end{equation}
where, from (\ref{fluc}), 
\begin{equation}
	\label{lamt}
		\rmi\hbar\frac{\partial}{\partial t}\Lk=\frac{\rmi\hbar}{\sqrt{\npk}}\left[\frac{\partial\hat{a}^{\dagger}_{c_{p(k)}}}{\partial t}\delta\hat{\Psi}_k(\bm{r})+\hat{a}^{\dagger}_{c_{p(k)}}\frac{\partial\delta\hat{\Psi}_k(\bm{r})}{\partial t}\right],
	\end{equation}
noting that the partial time derivative of $\npk$ is zero [as follows from (\ref{num_evol})] \cite{gm}. We find, straightforwardly [from (\ref{a}) and (\ref{delta}), respectively] that
\begin{subequations}
	\label{a_delta}
\begin{align}
	\rmi\hbar\frac{\partial\hat{a}^{\dagger}_{c_{p(k)}}}{\partial t}=&\sum_{k'=1}^{n}\delta_{k,k'}^p\int\left[\rmi\hbar\frac{\partial \phi_{k'}(\bm{r})}{\partial t}\right]\hat{\Psi}_{k'}^{\dagger}(\bm{r})\dr,\\
	\rmi\hbar\frac{\partial\delta\hat{\Psi}_k(\bm{r})}{\partial t}=&-\skd1n\Bigg\{\hat{a}_{c_{p(k')}}\int Q_{kk'}(\bm{r},\bm{r}')
	\left[\rmi\hbar\frac{\partial\phi_{k'}(\bm{r}')}{\partial t}\right]\dr'
	\nonumber\\&
	+\delta_{k,k'}^p\phi_k(\bm{r}) 
	\int\left[\rmi\hbar\frac{\partial\phi^*_{k'}(\bm{r}')}{\partial t}\right]\delta\hat{\Psi}_{k'}(\bm{r}')\dr'\Bigg\}.
\end{align}
\end{subequations}

\section{First order}
\label{app_1st}

The Hamiltonian, under a first-order approximation, is
\begin{widetext}
\begin{equation}
	\label{Ham1}
	\begin{split}
	\hat{H}_1=&\int\sk1n\Bigg\{\npk\phi_k^*(\bm{r})\left[H_{\text{sp}}^k(\bm{r})+\frac{\Utkk}{2}|\phi_k(\bm{r})|^2\right]\phi_k(\bm{r})
	+\sjk\Utjk\sqrt{\npj\npk}|\phi_j(\bm{r})|^2|\phi_k(\bm{r})|^2
	+\sum_{\substack{j=1\\j\neq k}}^n\npk\phi_j^*(\bm{r})H_{\text{ob}}^{jk}(\bm{r})\phi_k(\bm{r})\Bigg\}\dr
	\\
	&+\int\sk1n\Bigg(\sqrt{\npk}\bigg\{\phi_k^*(\bm{r})\bigg[H_{\text{sp}}^k(\bm{r})
	+\Utkk|\phi_k(\bm{r})|^2\bigg]\Lk+\text{H.c.}\bigg\}
	+\sjk\Utjk\bigg\{\sqrt{\npk}\left[\phi_j^*(\bm{r})\Lj+\text{H.c.}\right]|\phi_k(\bm{r})|^2
	\\&
	+\sqrt{\npj}\left[\phi_k^*(\bm{r})\Lk+\text{H.c.}\right]|\phi_j(\bm{r})|^2\bigg\}
	+\sum_{\substack{j=1\\j\neq k}}^n\sqrt{\npk}\bigg[\phi_j^*(\bm{r})H_{\text{ob}}^{jk}(\bm{r})\Lk
	+\Ljdag H_{\text{ob}}^{jk}(\bm{r})\phi_k(\bm{r})\bigg]\Bigg)\dr,
	\end{split}
\end{equation}
\end{widetext}
where we have assumed that 
\begin{subequations}
\begin{equation}
	\int\sk1n\npk\frac{\Utkk}{2}|\phi_k(\bm{r})|^4\dr\gg\int\sk1n\frac{\Utkk}{2}|\phi_k(\bm{r})|^4\dr,
\end{equation}
and
\begin{multline}
	\int\sjkn\sqrt{\npj\npk}\Utjk|\phi_j(\bm{r})|^2|\phi_k(\bm{r})|^2\dr \gg\\
	\int\sjkn\Utjk|\phi_j(\bm{r})|^2|\phi_k(\bm{r})|^2\dr,
\end{multline}
\end{subequations}
which, under the assumptions that the $N_c$ are large, is justified. In other words $\hat{H}_1=\hat{H}_3^{(N_c^1)}+\hat{H}_3^{(N_c^{1/2})}$.

To progress, we can now use the expression describing the evolution of the fluctuation operators (\ref{fluc_evol}), retaining terms up to first-order, to calculate
\begin{multline}
	\label{1storder}
	\rmi\hbar\frac{d\Lk}{dt}=[\Lk,\hat{H}_1]\\
	-\int\sum_{k'=1}^{n}\sqrt{\npk}Q_{kk'}(\bm{r},\bm{r}')\left[\rmi\hbar\frac{\partial\phi_{k'}(\bm{r}')}{\partial t}\right]\dr'.
\end{multline}
It is straightforward to obtain the expression for the commutator $[\Lk,\hat{H}_1]$: the terms linear in the fluctuation operator drop out immediately while the terms quadratic in the fluctuation operator can be written, in this first-order approximation, as $[\Lk,\Lkddd]= Q_{kk'}(\bm{r},\bm{r}')$ (\ref{fluccomzero}). Thus 
we can rewrite (\ref{1storder}) as
\begin{equation}
	\label{1storder_2}
	\begin{split}
	\rmi\hbar\frac{d\Lk}{dt}=&\int\sum_{k'=1}^n\sqrt{N_{c_{p(k)}}}\Bigg\{Q_{kk'}(\bm{r},\bm{r}')\Bigg[H_{\text{sp}}^{k'}(\bm{r}')\\
	&+{\tilde{U}_{k'k'}}|\phi_{k'}(\bm{r}')|^2-\rmi\hbar\frac{\partial}{\partial t}\\
	&+\sum_{\substack{j=1\\j\neq k'}}^n\tilde{U}_{jk'}\sqrt{\frac{N_{c_{p(j)}}}{{N_{c_{p(k')}}}}}|\phi_j(\bm{r'})|^2\Bigg]\phi_{k'}(\bm{r'})\\
	&+\sum_{\substack{j=1\\j\neq k'}}^n Q_{kj}(\bm{r},\bm{r}') H_{\text{ob}}^{jk'}(\bm{r}')\phi_{k'}(\bm{r}')\Bigg\}\dr'.
\end{split}
\end{equation}
We now note that the expectation value of the time derivative of the fluctuation operators is zero, i.e.\ $\langle d\tilde{\Lambda}(\bm{r})/dt\rangle =d\langle \tilde{\Lambda}(\bm{r})\rangle/dt=0$, so taking the expectation value of (\ref{1storder_2}) gives us the set of time-dependent Gross--Pitaevskii equations quoted in the main text.



\begin{thebibliography}{99}
\bibitem{foot1}
An extension of the formalism presented here which allows for spin-changing collisions, thus permitting the study of spinor condensates \cite{ueda}, is of course possible.
\bibitem{foot2}
Other features can be incorporated into the single-particle Hamiltonian: for example, one can explicitly include an externally applied rotation. In the rotating frame the single-particle Hamiltonian then becomes $\Hk\rightarrow\Hk+H_{\text{rot}}^k(\bm{r},t)$, where $H_{\text{rot}}^k(\bm{r},t)=\bm{\Omega}^{\text{rot}}_k(t)\cdot\bm{L}(\bm{r})$ is the component of the Hamiltonian detailing rotation. Here we have the rotation vector (in general specified for each component) giving the (possibly time dependent) rotation frequencies for each coordinate,  $\bm{\Omega}^{\text{rot}}_k(t)=\left(\Omega_k^{\text{rot}(x)}(t),\Omega_k^{\text{rot}(y)}(t),\Omega_k^{\text{rot}(z)}(t)\right)$ and angular momentum vector $\bm{L}(\bm{r})$ containing the angular momentum expressions for each coordinate, $\bm{L}(\bm{r})=\bm{r}\times[-\rmi\hbar\nabla]$.
\bibitem{foot3}
Note that in our Hamiltonian (\ref{Ham}), the one-body term $H_{\text{ob}}^{jk}(\bm{r},t)$ does not need to contain such a Kroneker delta mapping term since this is implicitly accounted for through the elements of $\Omega$.
\bibitem{foot4}
If one were to consider a traditional symmetry breaking approach, then the wave function should be partitioned as 
$\Pk=\Psi_k(\bm{r},t)+\delta\hat{\Psi}^{sb}_k(\bm{r},t)$, where $\Psi_k(\bm{r},t)=\left<\Pk\right>$ is a finite expectation value for component $k$ and $\delta\hat{\Psi}^{sb}_k(\bm{r},t)$ is a non-condensate (fluctuation) term for component $k$.
\bibitem{foot5}
Note that (\ref{qqq}) follows directly from the definition of $\delta\hat{\Psi}_k(\bm{r},t)$ (\ref{delta}) and the projector $Q_{kk'}(\bm{r},\bm{r}',t)$ (\ref{qdeq}). In the derivation though, one must be mindful of the need to apportion the correct Kronecker delta mapping term $\delta_{k,k'}^p$ (and its indexed variants).
\end{thebibliography}
\end{document}